\documentclass{article}
\usepackage{sao2,psfig,epsf}

\issue{2008, 63, 216-227}

\def\healpix{H{\sc ealpix }}
\def\glesp{G{\sc lesp }}

\def\wmap{\hbox{\sl WMAP~}}

\def\etal{et al.}
\def\alm{a_{\ell m}}
\def\Ylm{Y_{\ell m}}

\def\summ{\sum_{m=-\ell}^{\ell}}
\def\suml{\sum_{\ell=0}^{\infty}}

\def\le{\leq}

\begin{document}
\markboth{Naselsky, Verkhodanov, Nielsen}
	 {Instability of reconstruction of the low CMB multipoles}
\title{Instability of reconstruction of the low CMB multipoles}
\author{ P.D. Naselsky\inst{a}
    \and O.V. Verkhodanov\inst{b}
    \and M.T.B Nielsen\inst{a}
}
\institute{
Niels Bohr Institute, Blegdamsvej 17, DK-2100 Copenhagen, Denmark
\and \saoname }
\date{July 18, 2008}{May 17, 2008}
\maketitle

\begin{abstract}
We discuss the problem of the bias of the Internal Linear Combination (ILC)
CMB
map and show that it is closely related to the coefficient of
cross-correlation $K(\ell)$
of the true CMB and the foreground for each multipole $\ell$.
We present analysis of  the cross-correlation for the WMAP ILC quadrupole
and octupole from the
first (ILC(I)) and the third (ILC(III)) year data releases and
  show that these correlations are
$\sim -0.52-0.6$.   Analysing $10^4$ Monte Carlo simulations of the
random Gaussian CMB signals, we show that the distribution function for the
 corresponding coefficient of the cross-correlation has a polynomial shape
$P(K,\ell)\propto (1-K^2)^{\ell-1}$.
 We show that the most probable value of the
cross-correlation coefficient of the ILC and foreground quadrupole
has two extrema at $K\simeq\pm 0.58$.  Thus, the ILC(III) quadrupole
represents the
most probable value of the coefficient $K$.
We analyze the problem of  debiasing of the ILC CMB and pointed out  that
 reconstruction of the bias seems to be very problematic due to
statistical uncertainties. In addition, instability of the debiasing
illuminates itself
for the quadrupole and octupole components through the flip-effect, when the
even $\ell+m$ modes can be reconstructed  with significant error.
This error manifests itself
as opposite, in respect to the true sign of even low multipole modes,
and leads to significant
changes of the coefficient of cross-correlation with the foreground.
We show that
 the CMB realizations, whose
the sign of quadrupole $(2,0)$ component
is negative (and the same, as for  all the foregrounds),
the corresponding probability to get the positive sign after implementation
of the
ILC method  is about $40\%$.
\end{abstract}

\begin{keywords}
 cosmology: cosmic microwave background --- cosmology:
observations --- methods: data analysis
\end{keywords}

\section{INTRODUCTION}

Since ~~~ the~~~ COBE~~~ experiment~~~ and~~~ then~~ ~after~~ the~
Wilkinson~~~~ Microwave~~~ Anisotropy Probe\\ \mbox{ (WMAP)
\cite{wmapresults:Verkhodanov_n,wmapfg:Verkhodanov_n,wmapsys:Verkhodanov_n,wmap3temp:Verkhodanov_n,
wmap3ycos:Verkhodanov_n,wmap5temp:Verkhodanov_n,wmap5ycos:Verkhodanov_n}}
first and third year data releases, the problem of reconstruction
of the CMB low multipoles $\ell \le 10$ (including the  power of a
quadrupole, planarity and alignment of the quadrupole and the
octupole etc.) attracts very serious attention. A lot of
cosmological models have been involved for the explanation of the
peculiarities of the ILC (Internal Linear Combination )
low multipoles including the model of
running spectral index of the primordial
 adiabatic perturbations
\cite{wmapresults:Verkhodanov_n,wmapfg:Verkhodanov_n,wmapsys:Verkhodanov_n,wmap3temp:Verkhodanov_n,wmap3ycos:Verkhodanov_n},
non-trivial topology of the Universe
\cite{LuminetSuppressCMB:Verkhodanov_n,HornTopology:Verkhodanov_n},
Broken Scale Invariance (BSI) of the power spectrum
\cite{Bridges:Verkhodanov_n,Bean:Verkhodanov_n}, primordial
magnetic \mbox{field
\cite{kah:Verkhodanov_n,Durrer:Verkhodanov_n,ns04:Verkhodanov_n},}
Bianchi VIIh cosmological \mbox{model
\cite{jaffebian1:Verkhodanov_n,jaffebian2:Verkhodanov_n,mcewen:Verkhodanov_n,mcewenwmap3:Verkhodanov_n,gordon:Verkhodanov_n},}
etc.

The problem of low power of the CMB quadrupole can be related to
the non-Gaussianity of the WMAP low multipoles, including
planarity and alignment of the quadrupole and octupole
\cite{copi:Verkhodanov_n,copi3y:Verkhodanov_n,schwarz:Verkhodanov_n,evil:Verkhodanov_n}, cross--correlations with the
foregrounds \cite{ndv03:Verkhodanov_n,eriksenmf:Verkhodanov_n,fgxcorr:Verkhodanov_n,wmap3temp:Verkhodanov_n,saha:Verkhodanov_n}, local
defects of the CMB map \cite{cruz:Verkhodanov_n}, etc. These peculiarities need
some theoretical explanation, since the origin of those
non-Gaussian features still remains unknown.

Another way to look at the problem of the low multipole range of
the ILC CMB signal is to
 take under consideration the cross-correlation between the CMB signal
and the foregrounds, known as a bias of the ILC map
\cite{park3y:Verkhodanov_n,wmap3temp:Verkhodanov_n,saha:Verkhodanov_n,ndv03:Verkhodanov_n,ndv04:Verkhodanov_n}, some effects of
systematics \cite{pecquad:Verkhodanov_n} including contamination of the $\ell=2$
mode by dipole through far sidelobes \cite{dipolestraylight:Verkhodanov_n}.

   In this paper  we would like to re-examine one of the most important
issues of the data analysis of the modern CMB
experiments related to the problem of the CMB--foreground separation
by the ILC Method.
Needless to say, that separation of the CMB signal and foregrounds is
the major problem for all the CMB experiments
including the ongoing WMAP and the future Planck missions.
    The basic formalizm of the ILC method which allows one
to reconstruct the CMB signal from the observational data sets is
to combine the data in one (Linear Combination) map by using real
weighting coefficients. Then, to find these weighting
coefficients, we need to minimize some functional defined on
multifrequency CMB data sets containing information about the
primordial CMB signal, the instrumental noise and the foregrounds.
This functional (particularly, variance by Tegmark and Efstathiou
\cite{te:Verkhodanov_n} (hereafter TE)) can be defined both in the pixels domain
(like it was done by the WMAP team), and in the spherical
harmonics domain
\cite{toh:Verkhodanov_n} (hereafter TOH), and \cite{wmap3temp:Verkhodanov_n,ndv04:Verkhodanov_n,hansen:Verkhodanov_n}.
Below, we will discuss only the low CMB multipoles ($\ell \leq
10$) and, particularly, the quadrupole for which the ILC approach
in multipole domain seems to be very useful.


However,
the ILC approach in combination with the Minimal Variance Method
(hereafter, ILC\,MVM)
 produces a  specific feature of the
derived CMB signal, known as the bias. The bias reflects directly the
cross-correlation with  foregrounds,
and leads to the non-Gaussianity and statistical
anisotropy of the derived CMB through coupling with the foreground.
This is because  the MVM was originally  designed
for separation of the Gaussian signals, of which the statistical properties
are completely specified by the variance of each component.
At the same time,
it is well known
in the data analysis
that
the MVM is not motivated properly, especially in cases when we have
only a unique realization of the random process, where the CMB signal is
mixed with the significantly non-Gaussian foreground. Moreover,
the statistics of the low multipoles of the CMB signal are extremely
peculiar, since we do not have the statistical
ensemble of realizations.

In general, this paper is devoted to investigation of the ILC bias and
its statistical uncertainties. We will show that so-called debiasing of
the ILC CMB
has potential instabilities and unpredictability for a single realization
of the CMB.

  The outline of the paper is as follows.
In \mbox{Section 2}, we focus our attention on mathematical basis
of the ILC\,MVM
 exactly in the same way as it was proposed by \cite{wmap3temp:Verkhodanov_n}.
By analysis of the model with homogeneous foreground spectra, we
show that the bias of the ILC image, and  the bias of the power
spectrum are in order of $K(\ell)$ and $K^2(\ell)$, where
$K(\ell)$ is the coefficient of ``the true CMB --- true foreground''
cross-correlation. Following \cite{fgxcorr:Verkhodanov_n}, we will
call this cross-correlation ``the cosmic covariance''. Because of
statistical origin of this correlation, the cosmic covariance is
uncertain for each single realization of the CMB signal, even for
well known foregrounds. At the same time, derived by the ILC MVM,
 the ILC CMB signal and
corresponding ILC MVM foregrounds
have zero cross--correlations due to design of
the ILC \,MVM.

In Section 3, we draw our attention
to the investigation of the statistical properties of the   coefficient of
cross-correlation between the random Gaussian CMB and
the WMAP  foregrounds
for K--W bands. We generate  $10^4$  realizations of the Gaussian
CMB signals for the WMAP
best fit $\Lambda$CDM cosmological model, and cross-correlate them with
the foregrounds obtained by the WMAP team (the Maximum Entropy Method (MEM)
foregrounds, which constitute as a sum of the synchrotron, free-free
and dust emission for each frequency band). We will show,
that the probability distribution function for these cross-correlations
has a polynomial shape $P(K,\ell)\propto (1-K^2)^{\ell-1}$.
After that, we will find the most probable values of the coefficients
$\overline K(\ell)$ and $\overline K^2(\ell)$. In another words,
these coefficients determine the most probable value of the bias
for the ILC CMB and the power spectrum.
Taking under consideration that
the maps ILC(III) of three years and ILC(V) of five years data
are already debiased,
as the WMAP team claimed, we check out the cross-correlation between
the ILC and the WMAP forground. After debiasing, these cross-correlations
should represent the possible value of the $K(\ell)$ for the true CMB
and the true foreground.
For  the WMAP K-W bands,
the corresponding coefficients of the cross--correlations are
about $K\simeq -0.51$. Remarkably, this
value is close to the most probable value  $\overline K(\ell)$
of the distribution function $KP(K,\ell=2)$.

In Section~4 we discuss some peculiarities of the ILC(III)
quadrupole and the octupole. As it follows from the Monte Carlo
simulations presented in \cite{eriksen2004a:Verkhodanov_n}
  for low multipoles $\ell=2,3$,
the sign of the ($\ell=$2,$m=$0), (2,2) and (3,1) components
  was changed to the opposite one
for about 20\% of the output CMB maps. This flip-effect
significantly changed the morphology of
the output CMB map.
We will show that the flip-effect takes place
for those input CMB quadrupoles,
for which the sign of $\ell=2,m=0$ mode  is opposite
to the sign of the foreground.
For the Gaussian CMB signal about
  50\% of the realizations have \mbox{${\rm sign}~c_{2,0}=-{\rm sign}~F_{2,0}$,}
where $c_{2,0}$ and $F_{2,0}$ are the $m$=0 components of the quadrupole
for the input CMB and the foreground, respectively.
Roughly for $40\%$ of those realizations, reconstruction of the
  \mbox{$\ell=2,m=0$} component could have a wrong sign of the (2,0) mode.
In this Section we present some analytical description of the flip-effect
based on the properties of the bias.

 We summarize our results in Conclusion.

\section{MATHEMATICAL BASIS OF THE ILC MVM METHOD}

Let us draw our attention to the general formalism of the ILC method,
assuming that we have experimental data
$\Delta T(\theta_p, \phi_p,\nu_i)$ for the CMB anisotropy
 obtained for some frequency bands $\nu_1, \nu_2...$.
Here $\theta_p$ and $\phi_p$ are the standard polar and
azimuthal angles of the spherical coordinate system and the index $p$
mark a corresponding pixel.
The observed sky temperature for each pixel $p$ in
a given band $i$ is the linear combination of the
true CMB signal $T_c(p)$ and the true foreground $f_i(p)$.
In reality, the components of the foreground at different frequency bands
(mainly the synchrotron, the free-free and the dust emission)
 have angular variations of the spectral indicies.
However, by using different disjoint regions of the
sky with nearly uniform foreground spectra and low level of spatial
variations we can simplify the
mathematical basis of the ILC method
\cite{wmap3temp:Verkhodanov_n}.  
Following  the WMAP model of the foregrounds, let us assume that
for some region of the sky the spatial variation of the
foregrounds emission are negligible and $f_i(p)=S_i F(p)$, where
$S_i$ is the frequency spectrum and $F(p)$ is the spatial
distribution  of the foreground for each band $i$. The main idea
of the ILC method is to estimate the CMB signal by using real
weighting coefficients $w_i$ for each region of the sky with
monotonic spectrum of the foreground \cite{wmap3temp:Verkhodanov_n}
\begin{equation}
 \small{ T_{ilc}(p)=\sum_iw_iT_i(p)=T_c(p)+\Gamma F(p),\hspace{0cm}\sum_i w_i=1,
\label{h1:Verkhodanov_n}}
\end{equation}
where $\Gamma=\sum_iw_iS_i$. The coefficients $w_i$ can be found
by minimization of the variance (see \cite{wmap3temp:Verkhodanov_n})
\begin{eqnarray}
\nonumber \sigma^2_{ilc}=\langle T^2_{ilc}(p)\rangle - \langle
T_{ilc}(p)\rangle^2\\=
     \sigma^2_c +2\Gamma \sigma_{cF}+\Gamma^2\sigma^2_F. \hspace{0.6cm}
\label{h2:Verkhodanov_n}
\end{eqnarray}
Here $\sigma^2_c$ is the variance of the true CMB, $\sigma^2_F$ is
the variance of the true foreground and
$$\sigma_{cF}=\langle T_c(p)F(p)\rangle -\langle T_c(p)\rangle\langle
F(p)\rangle$$ is the cosmic covariance between the true CMB and
the foreground
\cite{fgxcorr:Verkhodanov_n}. 
The angle brackets in Eq.\ref{h2:Verkhodanov_n} and hereafter
denote the averages over the pixels belonging to each zone of the
map with the monotonic spectrum of the foreground. The minimum of
the variance $\sigma^2_{ilc}$ can be easily found from the
equation $\frac{d\sigma^2_{ilc}}{dw_i}=0$, whence we get
\cite{wmap3temp:Verkhodanov_n}  
\begin{eqnarray}
 T_{ilc}(p)=T_c(p)-\frac{\sigma_{cF}}{\sigma^2_F}F(p),\hspace{0.1cm}
\sigma^2_{ilc}=\sigma^2_c-\frac{\sigma^2_{cF}}{\sigma^2_F}.
\label{h3:Verkhodanov_n}
\end{eqnarray}
Let us define the coefficient of cross-correlation between the
true CMB signal and the true \mbox{foreground as}
 \begin{eqnarray}
  \kappa=\frac{\sigma_{cF}}{\sigma_c\sigma_F}.
\label{h4:Verkhodanov_n}
 \end{eqnarray}
Then,~~ after~~ substitution ~~of~~ Eq.(\ref{h4:Verkhodanov_n})~~
in~~ Eq.(\ref{h3:Verkhodanov_n})~~ we \mbox{have
\cite{fgxcorr:Verkhodanov_n}:} 
\begin{eqnarray}
 T_{ilc}(p)=T_c(p)-\kappa\frac{\sigma_c}{\sigma_F}F(p),
            \hspace{0.1cm}\sigma^2_{ilc}=\sigma^2_c(1-\kappa^2).
\label{h5:Verkhodanov_n}
\end{eqnarray}
As one can see from Eq.(\ref{h5:Verkhodanov_n}), the temperature
per pixel $T_{ilc}(p)$ and the variance  $\sigma^2_{ilc}$ are
biased. For $T_{ilc}(p)$ the difference $T_{ilc}(p)-T_c(p)$ is
proportional to the $\kappa$ and can be both positive and
negative. For the variance $\sigma^2_{ilc}$, the bias is always
negative and proportional to $1-\kappa^2$. The existence of such a
bias is not supprising. The shift of the ILC variance and
corresponding power spectrum was discussed in
\cite{wmap3temp:Verkhodanov_n,park3y:Verkhodanov_n,saha:Verkhodanov_n,fgxcorr:Verkhodanov_n}.
However, it would be important to note that the bias of the variance
and the power spectrum is determined by the cosmic covariance,
which is unknown for each single realization of the CMB sky.
This is the reason why it can not be corrected by any methods
because of the statistical uncertainties.

\begin{table*}
\caption{The table of $K_{ff}(\ell)$ coefficients for ``V--W'', ``Ka--V'',
``Ka--W'' and ``Q--V'' MEM foregrounds. The first column indicates the
multipole $\ell$.
}
\begin{tabular}{c|c|c|c|c} 
\hline
$K_{ff}(\ell)$&$V-W$&$Ka-V $&$Ka-W $&$Q-V$\\ 
\hline
2 &0.999969 &0.999014 & 0.998761 &0.999481 \\
\hline
3&0.994699 &0.999534 &0.995187 &0.999770 \\
\hline
4&0.996702 &0.999497 &0.996098  &0.999708  \\
\hline
5 &0.994615 &0.998862  &0.995825 &0.999647 \\
\hline
6&0.995481 &0.999354 &0.993799 &0.999563 \\
\hline
7& 0.993245&0.998620 &0.995323 &0.999506  \\
\hline
8 &0.994281 &0.999252 & 0.993458 &0.999575 \\
\hline
9&0.993532 &0.998781 &0.994617 &0.999571 \\
\hline
10&0.994638 &0.999068 & 0.993319& 0.999463 \\
 \hline
\end{tabular}
\label{Kff:Verkhodanov_n}
\end{table*}
The ILC approach presented above in the pixel domain can be easily
generalized for the multipole domain by
 spherical harmonics decomposition:
$\Delta T(\theta, \phi,\nu_i)\rightarrow a^{\nu_i}_{\ell,m}$:
\begin{equation}
\Delta T(\theta,\varphi,\nu_j)=\suml \summ |a^{\nu_j}_{\ell,m}|
          e^{i\phi_{\ell,m}} \Ylm (\theta,\varphi),
\label{eq1:Verkhodanov_n}
\end{equation}
where $|\alm|$ and $\phi_{\ell,m}$ are the moduli and phases of
the coefficients of the expansion. Direct conversion of
Eq.(\ref{h5:Verkhodanov_n}) to the multipoles coefficients gives
us
\begin{eqnarray}
 \overline c^{ilc}_{\ell,m}=c_{\ell,m}-
        \kappa\frac{\sigma_c}{\sigma_F}F_{\ell,m}.
\label{tca7:Verkhodanov_n}
\end{eqnarray}
It is easy to show that this solution for the derived ILC signal
$\overline c^{ilc}_{\ell,m}$ can be obtained by minimization of the variance
\begin{eqnarray}
\sigma^2=\sum_\ell\frac{2\ell+1}{4\pi}\sum_{m=-\ell}^\ell|
      \sum_jw_ja^{j}_{\ell,m}|^2.
\label{tca8:Verkhodanov_n}
\end{eqnarray}
Moreover, follow to TOH
approach
   we can define the ILC CMB in multipole domain
by using weighting coefficients $W_i(\ell),\sum_iW_i(\ell)=1$
and minimizing  the power spectrum for each multipole
$$C(\ell)=\sum_m|W_i(\ell)a^{(i)}_{\ell,m}|^2.$$
For the model of uniform spectra of the foreground this approach
leads to
\begin{eqnarray}
 c^{ilc}_{\ell,m}=c_{\ell,m} -K(\ell)\frac{(\sum_{m=-\ell}^\ell
|\overline c_{\ell,m}|^2)^{\frac{1}{2}}}
{\left(\sum_{m=-\ell}^l|F^{(2)}_{\ell,m}|^2\right)^{\frac{1}{2}}}
           F^{(2)}_{\ell,m},
\label{tt9:Verkhodanov_n}
\end{eqnarray}
where $K(\ell)$ is
the coefficient of cross-correlation
 between the ILC CMB and the foreground:
\begin{eqnarray}
K(\ell)=\frac{\sum_m\left(c^{ilc}_{\ell,m}F^{*}_{\ell,m}+
       (c^{ilc})^*_{\ell,m}F_{\ell,m}\right)}
       {2\left[\sum_m |c^{ilc}_{\ell,m}|^2
       \sum_m|F_{\ell,m}|^2\right]^{\frac{1}{2}}}.
\label{tt10:Verkhodanov_n}
\end{eqnarray}
Taking Eq.(\ref{tt9:Verkhodanov_n}) and
Eq.(\ref{tt10:Verkhodanov_n}) into account, one can get
\begin{eqnarray}
C^{ilc}(\ell)=C_c(\ell)\left[1-K^2(\ell)\right].
\label{tt11:Verkhodanov_n}
\end{eqnarray}

Before any future analysis, let us draw our attention to the
applicability of the model of the uniform foreground spectra,
taking the MEM foregrounds in K--W bands of the WMAP data
into account.
For these
foregrounds (sum of the synchrotron, free-free and dust emission
for each frequency band) the coefficient of cross-correlation is
defined in the same way as Eq.(\ref{tt10:Verkhodanov_n}) with
substitution of $F^{(j)}_{\ell,m},F^{(k)}_{\ell,m}$ for
$c^{ilc}_{\ell,m},F_{\ell,m}$ in Eq.(\ref{tt10:Verkhodanov_n}).

\begin{figure*}
\hbox{\hspace*{-0.5cm}
\centerline{
\psfig{figure=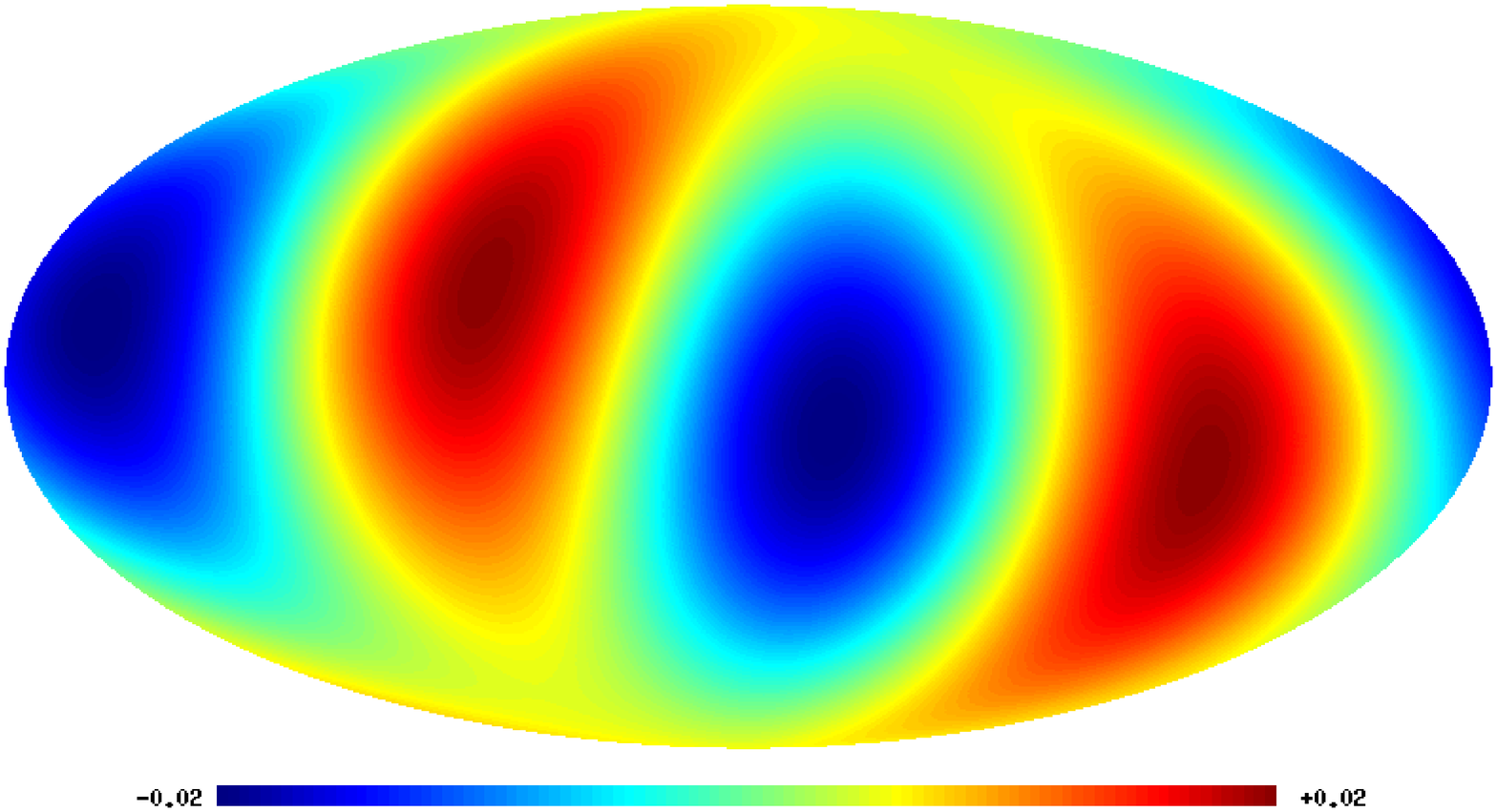,width=6cm}
\psfig{figure=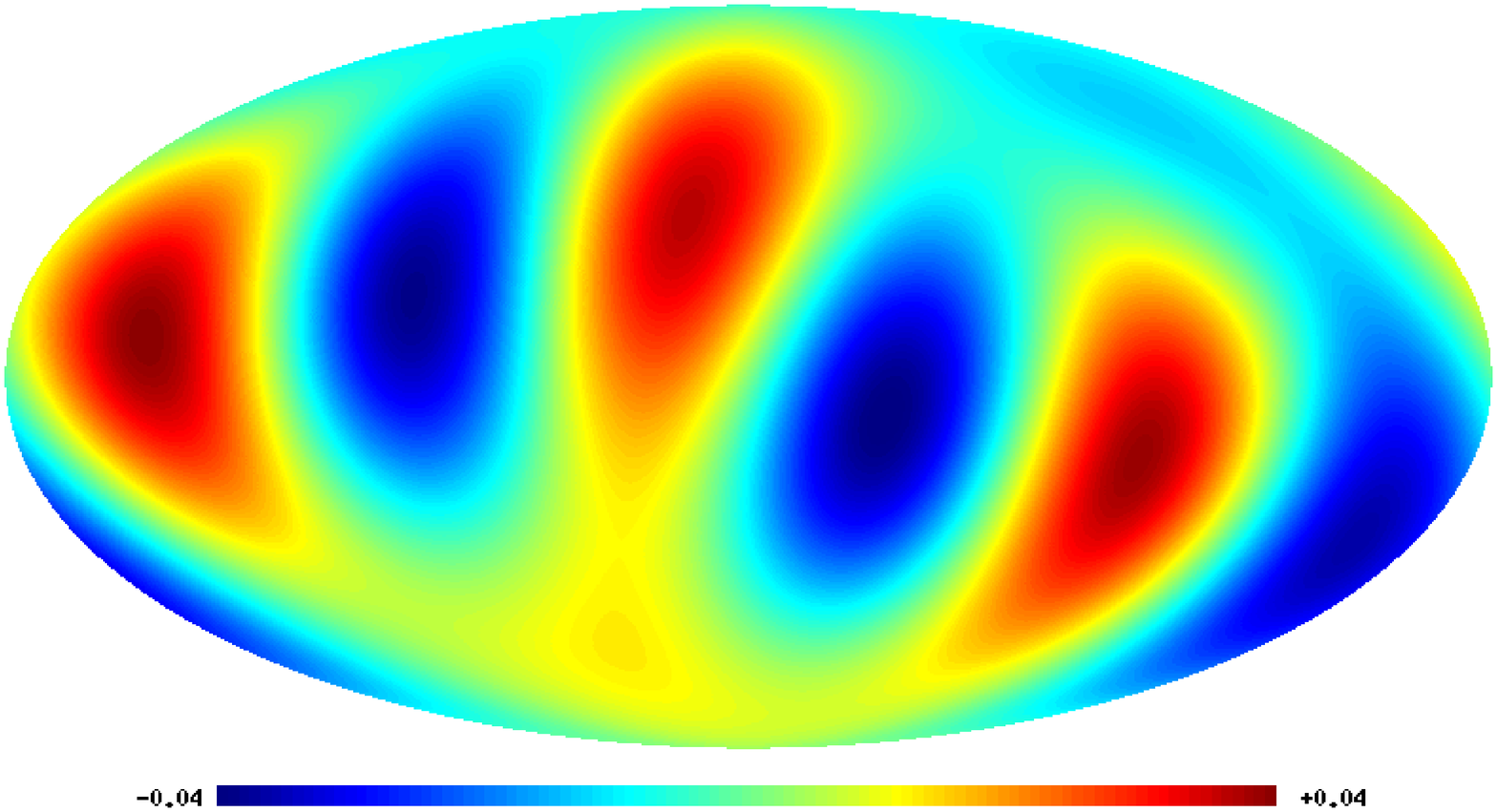,width=6cm}
}}
\hbox{\hspace*{-0.5cm}
\centerline{
\psfig{figure=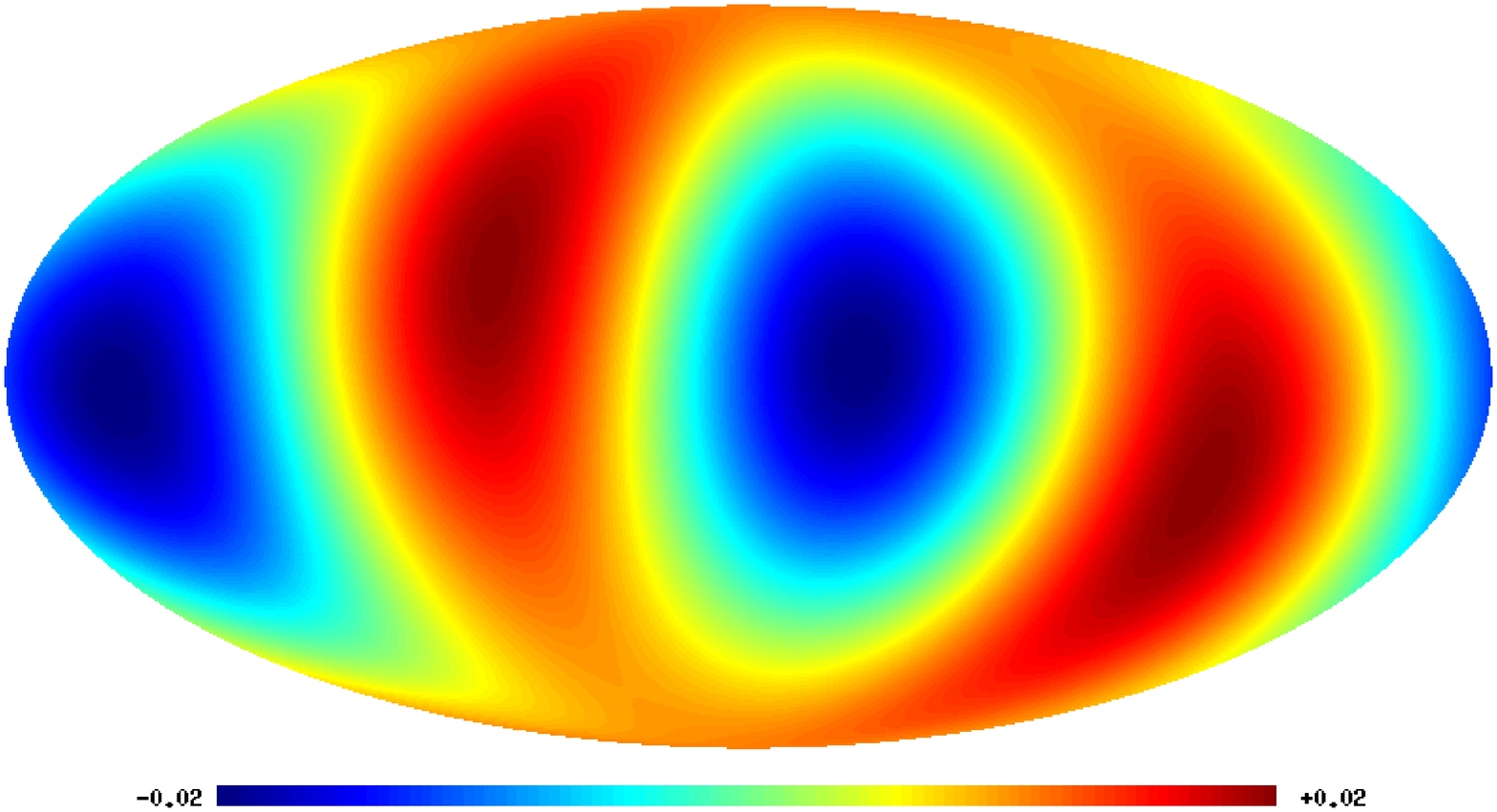,width=6cm}
\psfig{figure=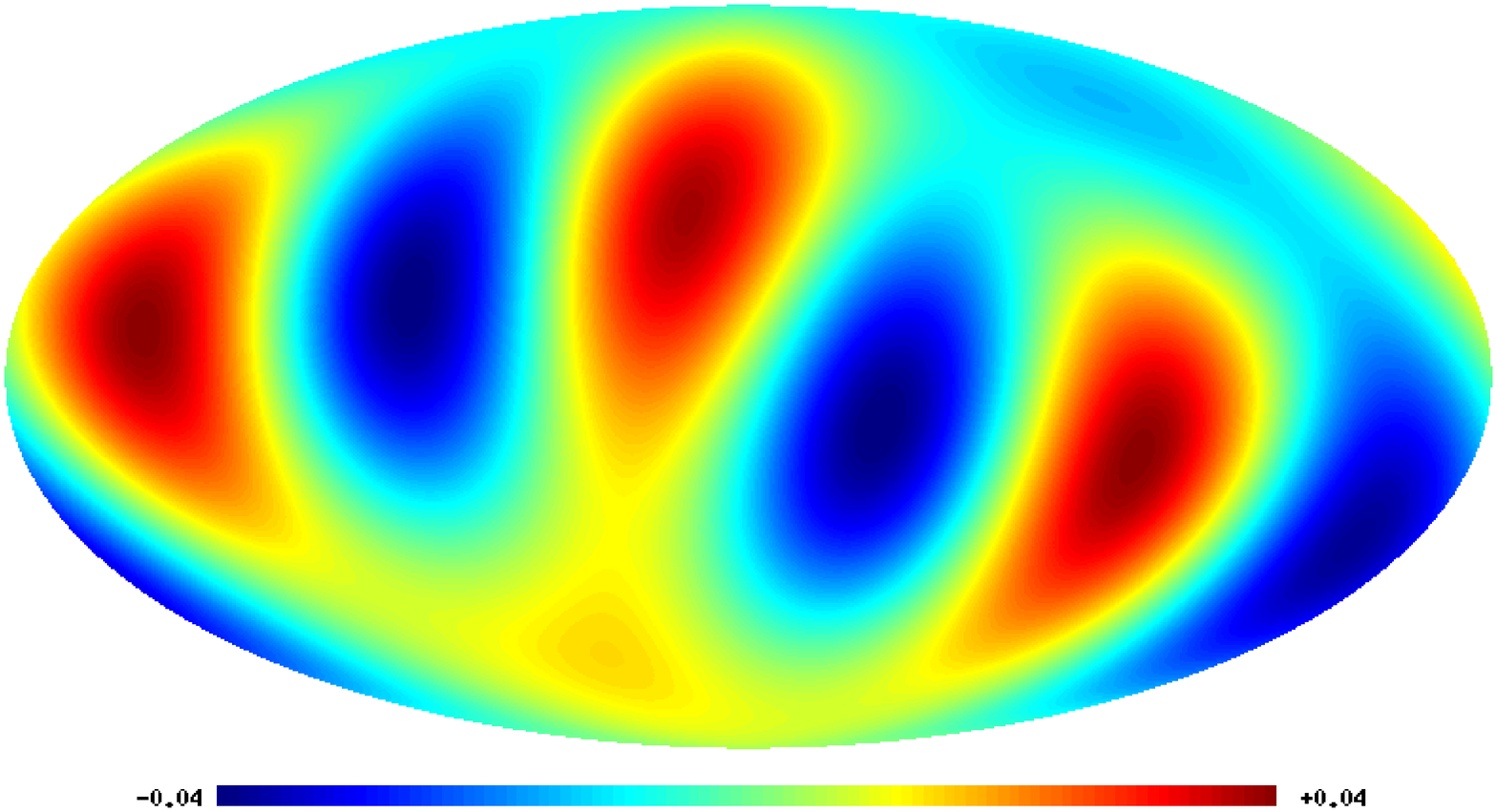,width=6cm}
}}
\hbox{\hspace*{-0.5cm}
\centerline{
\psfig{figure=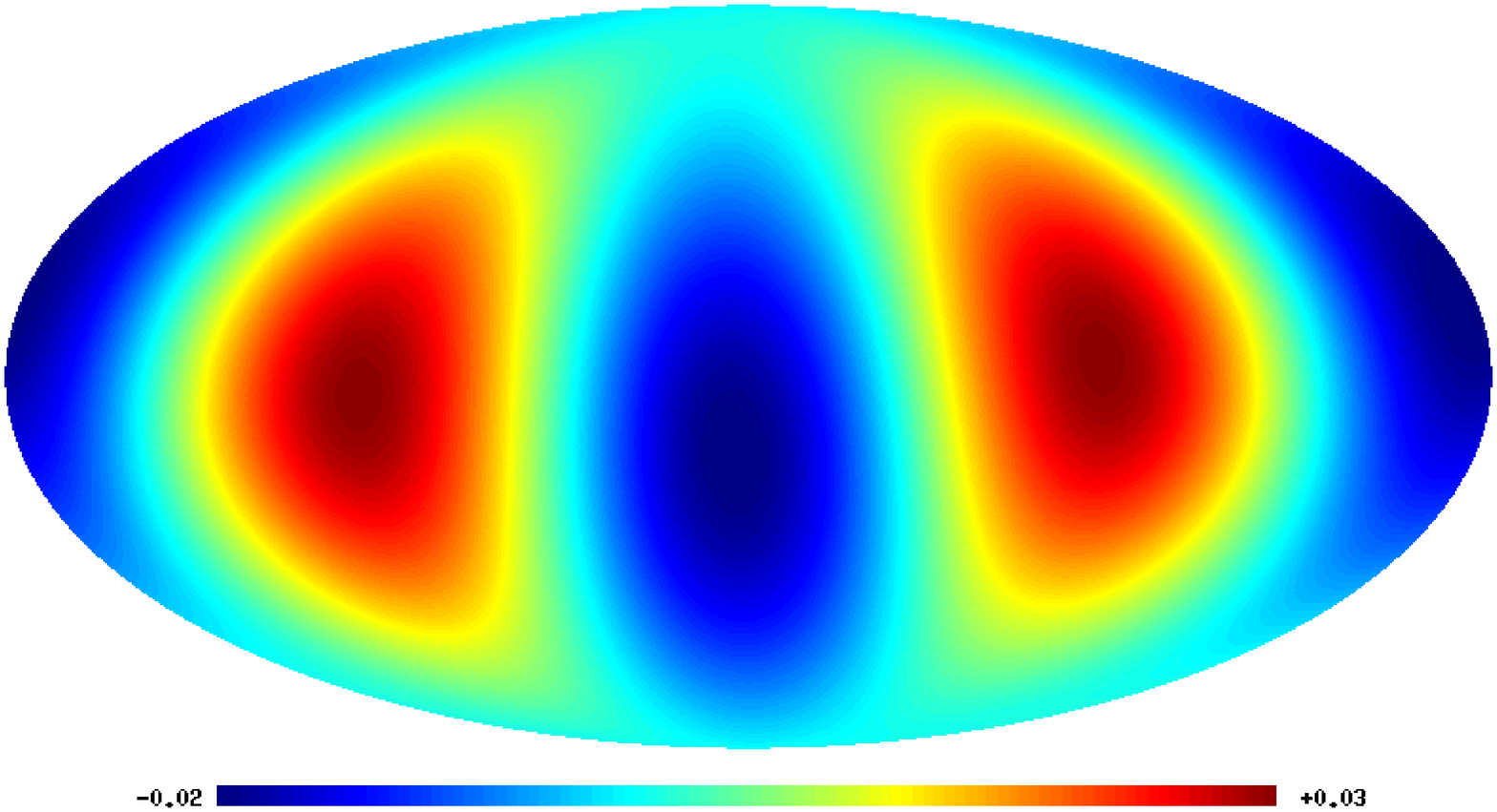,width=6cm}
\psfig{figure=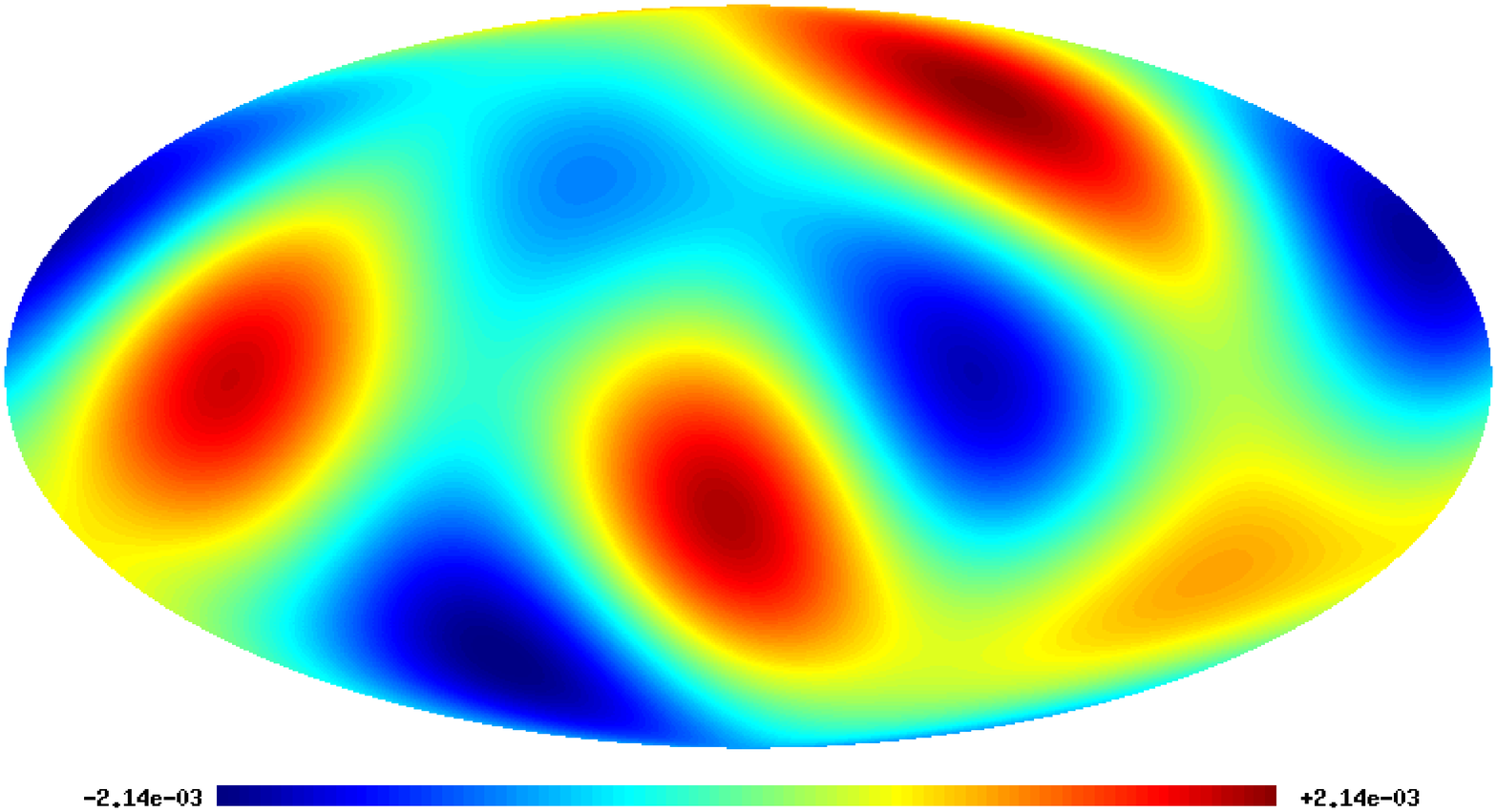,width=6cm}
}}
\caption{The quadrupole component of the ILC(I) (top left), the
ILC(III) quadrupole (middle left) and the difference between them
(bottom left). Right column  is for the ILC(I), ILC(III) and
difference of octupole. }
\label{fb1:Verkhodanov_n}
\end{figure*}

In Table\,\ref{Kff:Verkhodanov_n} we show the coefficients of
cross-correlation for different combination of the WMAP MEM
foregrounds. One can see that corresponding coefficients of
cross-correlation between different foregrounds (for different
bands) are extremely close to unity. Thus the model of uniform
spectra of the foreground is very well motivated. As it follows
from Eq.(\ref{tt9:Verkhodanov_n}) and
Eq.(\ref{tt11:Verkhodanov_n}), the corresponding bias of the ILC
coefficient $c^{ilc}_{\ell,m}$ is proportional to the $K(\ell)$,
while the bias of the power spectrum is of the order of $K^2(\ell)$.
Since the true CMB constitute as a random (Gaussian ?) process,
the coefficient of cross-correlation $K(\ell)$ represents a random
(non-Gaussian) process as well. Thus, for
 each particular realization of the CMB sky, the coefficient of
cross-correlation $K(\ell)$ remains uncertain.
Note that the WMAP team, as mentioned in
\cite{saha:Verkhodanov_n}, to correct the CMB power
obtained by the ILC method took under consideration $10^2\div10^4$
realizations of the random Gaussian CMB, creating the statistical
ensemble of realizations. However, even after the averaging over
realizations  we can not predict the bias for one particular
realization of the CMB due to the cosmic covariance.
Moreover, as it is seen from Eqs.(\ref{tt9:Verkhodanov_n}) and
(\ref{tt11:Verkhodanov_n}), the average of the ILC CMB signal
over the ensamble of realization  reduces the bias factor $\langle
K \rangle$ down to zero, while  $\langle K ^2\rangle\neq 0 $.
Thus, for the ensemble of realization the correction of the CMB
power can be done successfully, while the correction of the signal
itself seems to be very problematic.

One more problem is related to the statistical properties of
the true CMB signal. As it is well known, the low multipoles ($\ell\le 10$)
of the WMAP CMB signal reveal significant departure from
the statistical isotropy and homogenity
\cite{eriksen2004a:Verkhodanov_n,eriksen3yasym:Verkhodanov_n}. 
If this effect is the manifestation of the  primordial non-Gaussianity of
the CMB, we simply do not have the correct model of the statistical ensemble
of realizations.
This remark illuminates
significantly  stronger
the problem of
instability of the reconstruction of the low multipoles of the CMB signal.

Not to be able to give  the general solution of the problems
mentioned above, we would like to propose indirect method of
evaluation of the bias in the image and power domains by looking at
the cross-correlation coefficient $K(\ell)$ of the derived  CMB
signal and the MEM  foregrounds, including the
the synchrotron map of \mbox{Haslam et al.}
\cite{haslam:Verkhodanov_n} (hereafter HA).
The idea is  based on the fact that
 in Eq.(\ref{h5:Verkhodanov_n}) the cross-correlation of the ILC and the foreground
is exactly zero. Thus, if by implementation some methods of debiasing
of the ILC map we will get $K\neq 0$, it would be reasonable accurate
estimator of the initial  coefficient of cosmic covariance $\kappa$.  In
 next section we will discuss this approach in details.

\section{THE ILC(I) AND ILC(III) CROSS-CORRELATIONS WITH THE FOREGROUNDS}

   As was pointed out by the WMAP team
\cite{wmap3temp:Verkhodanov_n}  
the correction
of the ILC CMB can be done by implementation of additional information about
the foregrounds and
 100 realizations of the random Gaussian CMB signals, which mimic
the WMAP K-W bands
in combination with MEM foreground.
 Repeating the ILC MVM method for each  from these 100 realization,
the WMAP team took under consideration some systematic of the ILC
method and made corresponding
correction of the derived the ILC(III)  map.

\begin{figure*}
\hbox{\hspace*{-0.5cm}
\centerline{\psfig{figure=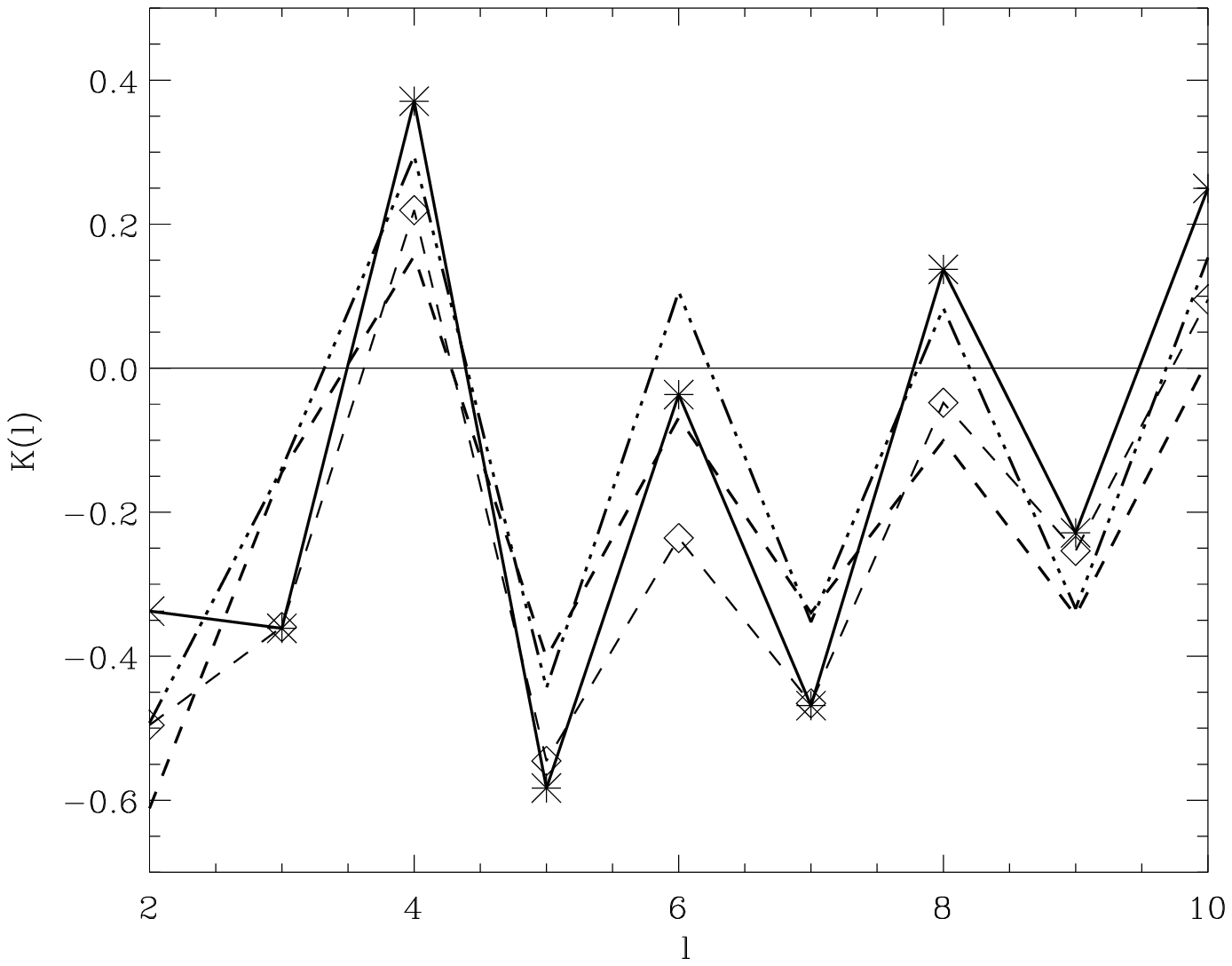,width=7cm}}}
\hbox{\hspace*{-0.5cm}
\centerline{\psfig{figure=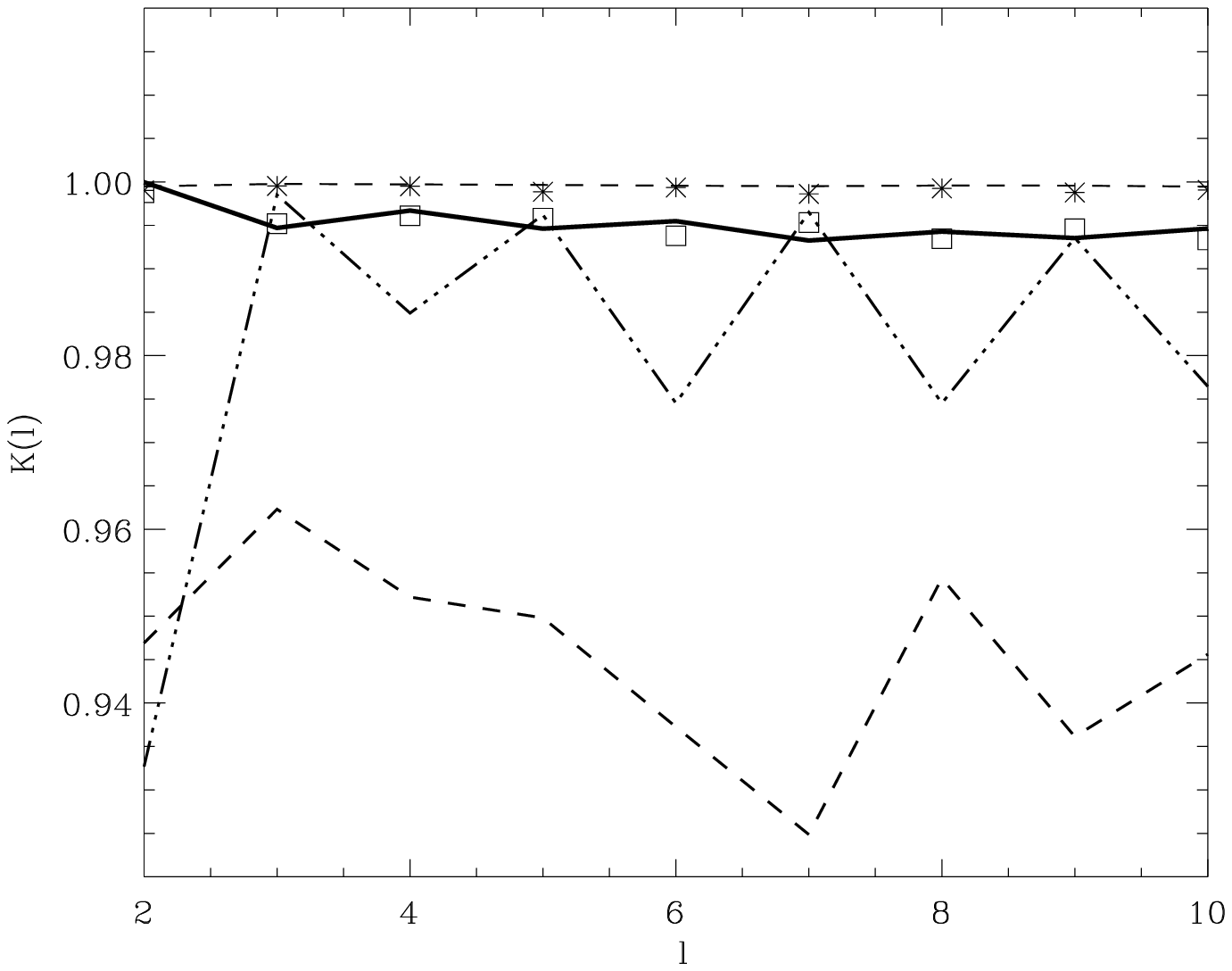,width=7cm}}}
\caption{Top.
The coefficient of cross--correlation of the ILC(I) and ILC(III) with the
foregrounds.
The dash  line is for the ILC (III) and the Haslam et al. map
(1982);
the triple dots dash line is for the ILC(I) and Haslam et al. map;
the thick solid line with stars
is  for the cross-correlation of ILC(I) and V-band MEM foreground.
The dash line with diamonds is for the ILC(III) and
the V-band MEM foreground.
Bottom.
The coefficients of cross-correlations between V- and
W-foregrounds of the WMAP (the thick solid line),
V- and Haslam et al. 1982 (the dash line)
and finally ILC(I) and ILC(III) (the triple dots--dash line).
The dash line is for Q- and V-bands, the stars are for Ka--V and
the boxes are for Q--V foregrounds.
}
\label{kcross:Verkhodanov_n}
\end{figure*}


It is important to note that the WMAP team
\cite{wmap3temp:Verkhodanov_n}   
pointed out that difference between the ILC(III) and the ILC(I) is
mainly due to the bias. In Fig.\,\ref{fb1:Verkhodanov_n} we show the
quadrupole and the octupole components of the ILC(I) and the
ILC(III) maps and difference between them to illustrate the
correction of the bias made by the WMAP team for low multipoles
range of the signal.

Let us draw our attention to the cross-correlation of the ILC(I) and
ILC(III)  maps with the
MEM foregrounds (being the  sum over synchrotron, free--free and dust
emission maps).
 In Fig.\,\ref{kcross:Verkhodanov_n} we show the coefficient
of the ILC-foreground cross-correlation $K(\ell)$ for the
multipoles $\ell=2-10$ and for the WMAP MEM foreground for K--W
band\footnote{For the ILC(I) we use the WMAP MEM foregrounds from
the first year data release.}. In addition in
Fig.\,\ref{kcross:Verkhodanov_n}, we show  $K_j(\ell)$ for the
ILC(I), the ILC(III), and the Haslam et al. synchrotron map
\cite{haslam:Verkhodanov_n}. 
As one can see from
this figure, the quadrupole $(\ell=2)$ and $\ell=5$ multipole of the ILC(III)
have negative correlation ($K\sim -0.5$) with Ka--W  foregrounds.
Moreover, the ILC(III) quadrupole has $K\sim -0.6$
with the Haslam et al. quadrupole.

\subsection{Statistical properties of the bias}

To understand the properties of the  ILC(III)---foreground
cross-correlation, we performed numerical test taking under
consideration 10000 realizations of the random Gaussian CMB maps
and cross-correlate them with the same model of the foregrounds as
in  Fig.\,\ref{kcross:Verkhodanov_n} without any debiasing.

\begin{figure*}
\hbox{\hspace*{-0.5cm}
\centerline{
\psfig{figure=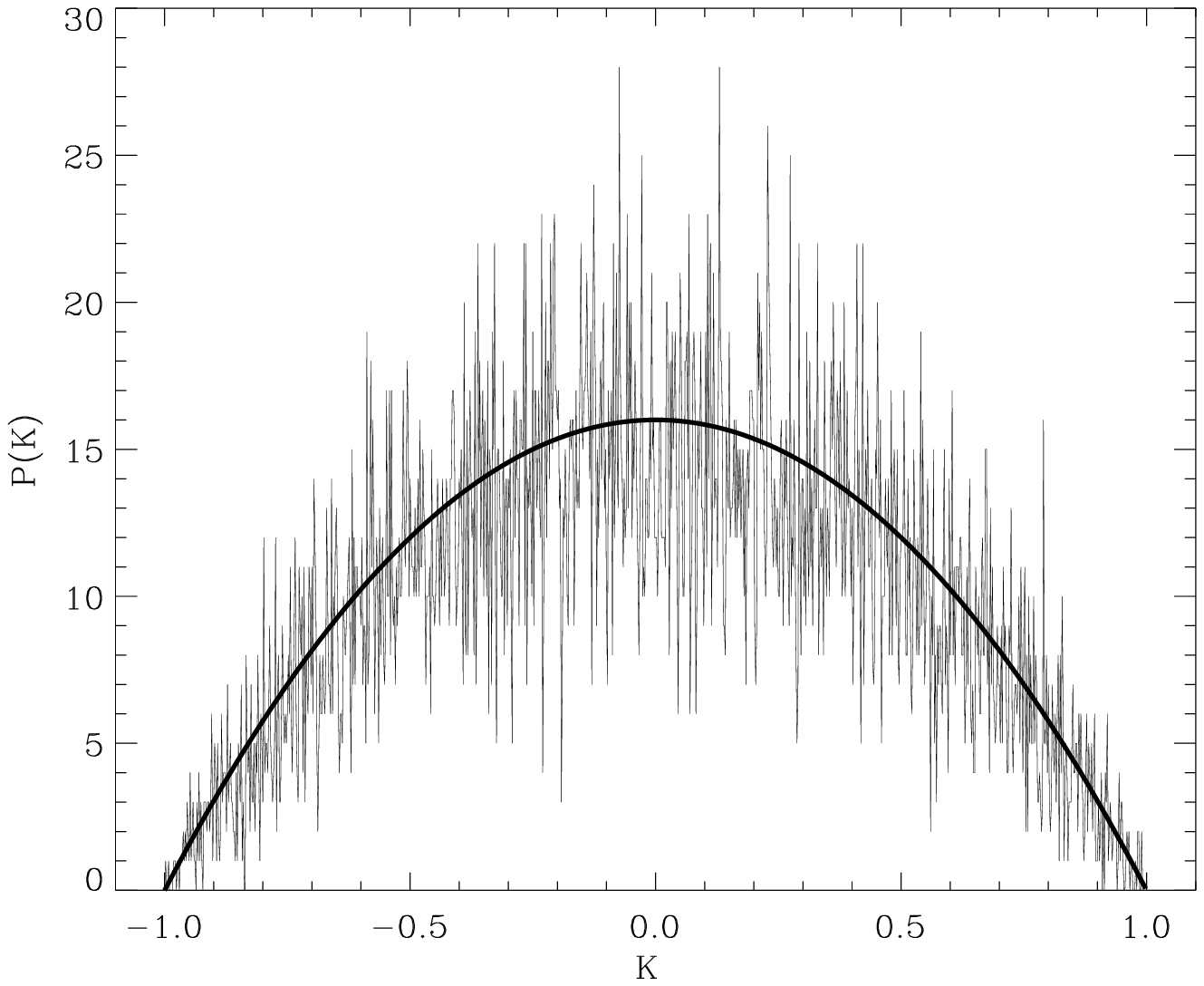,width=5cm}
\psfig{figure=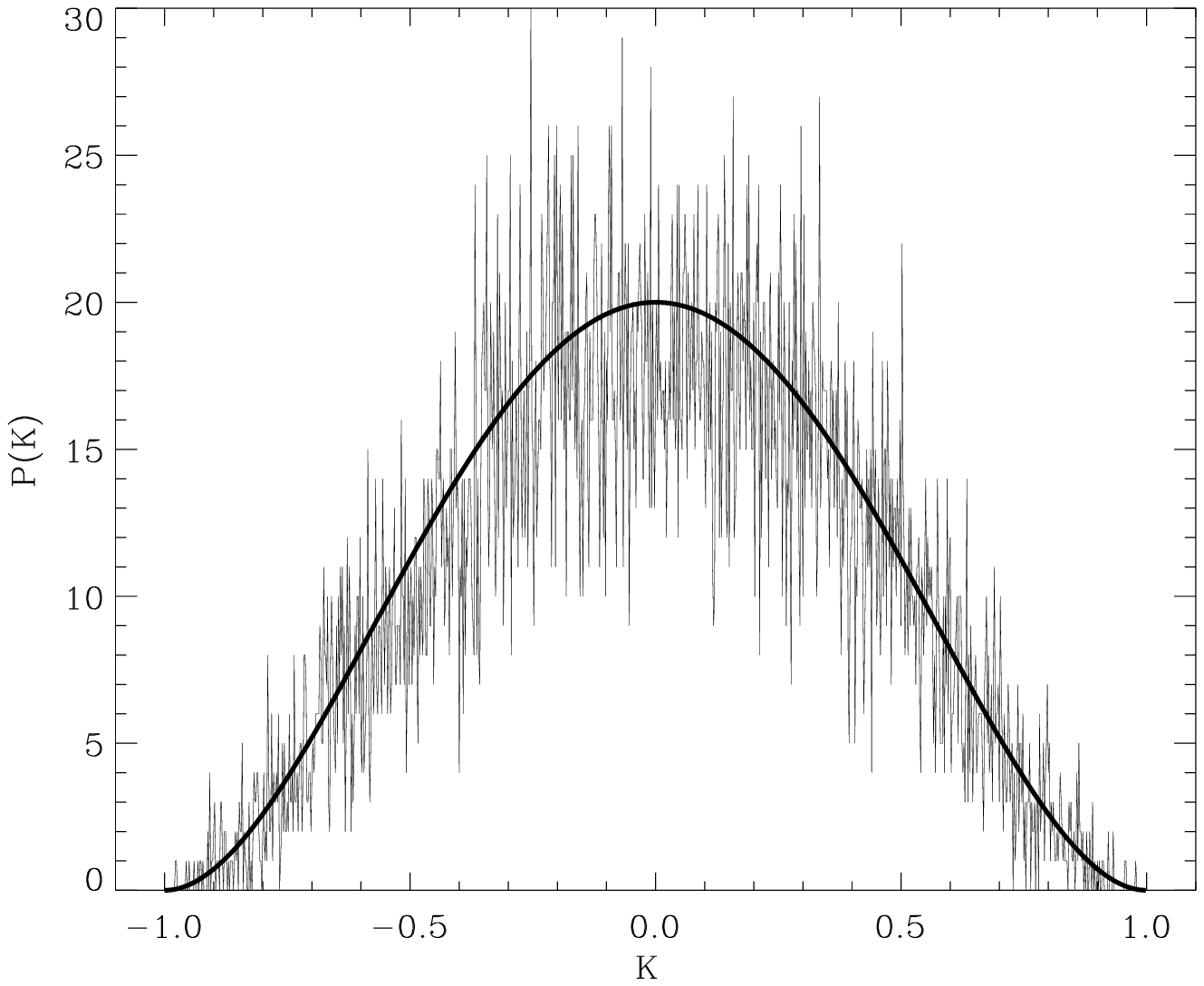,width=5cm}
\psfig{figure=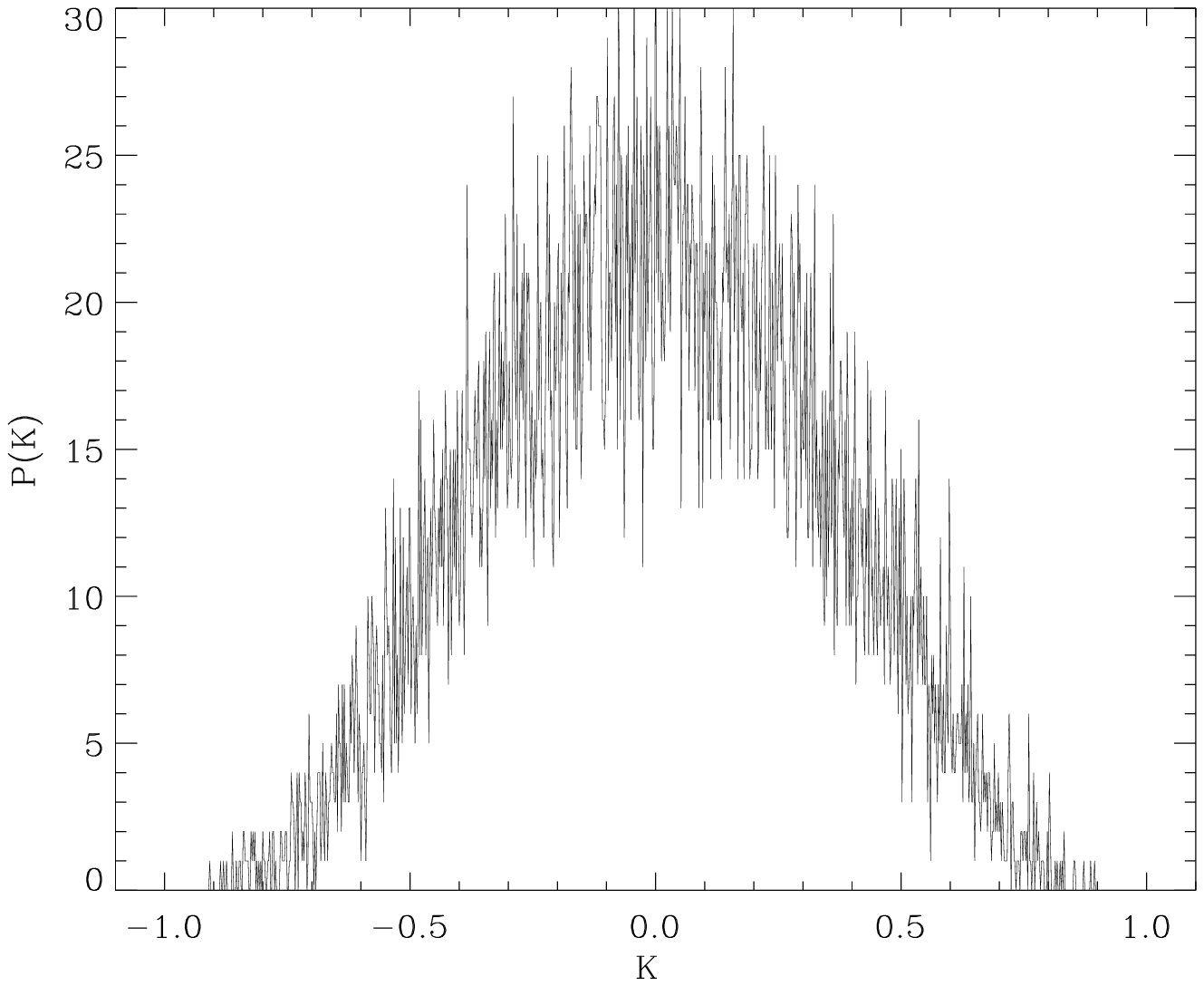,width=5cm}}}
\hbox{\hspace*{-0.5cm}
\centerline{
\psfig{figure=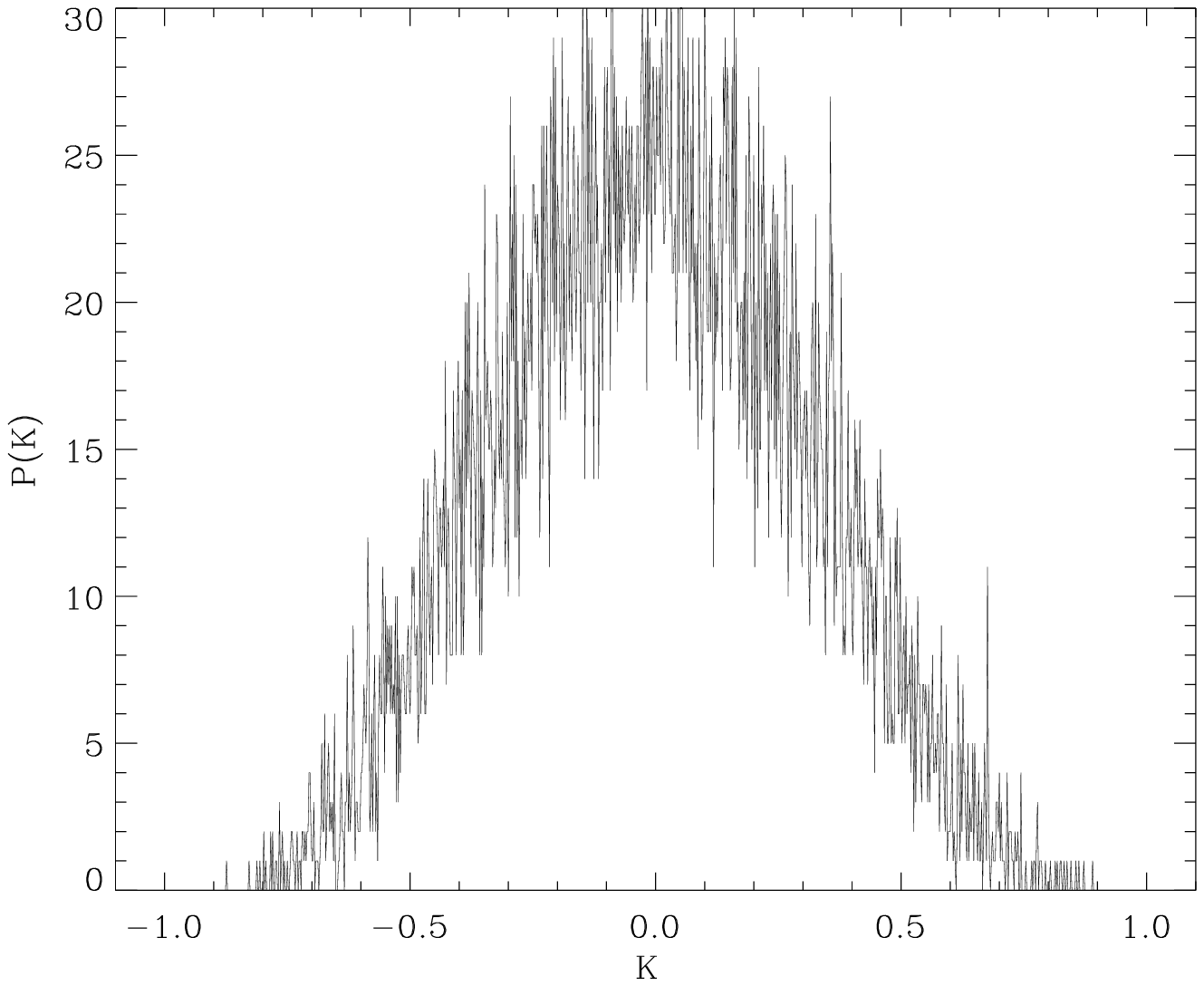,width=5cm}
\psfig{figure=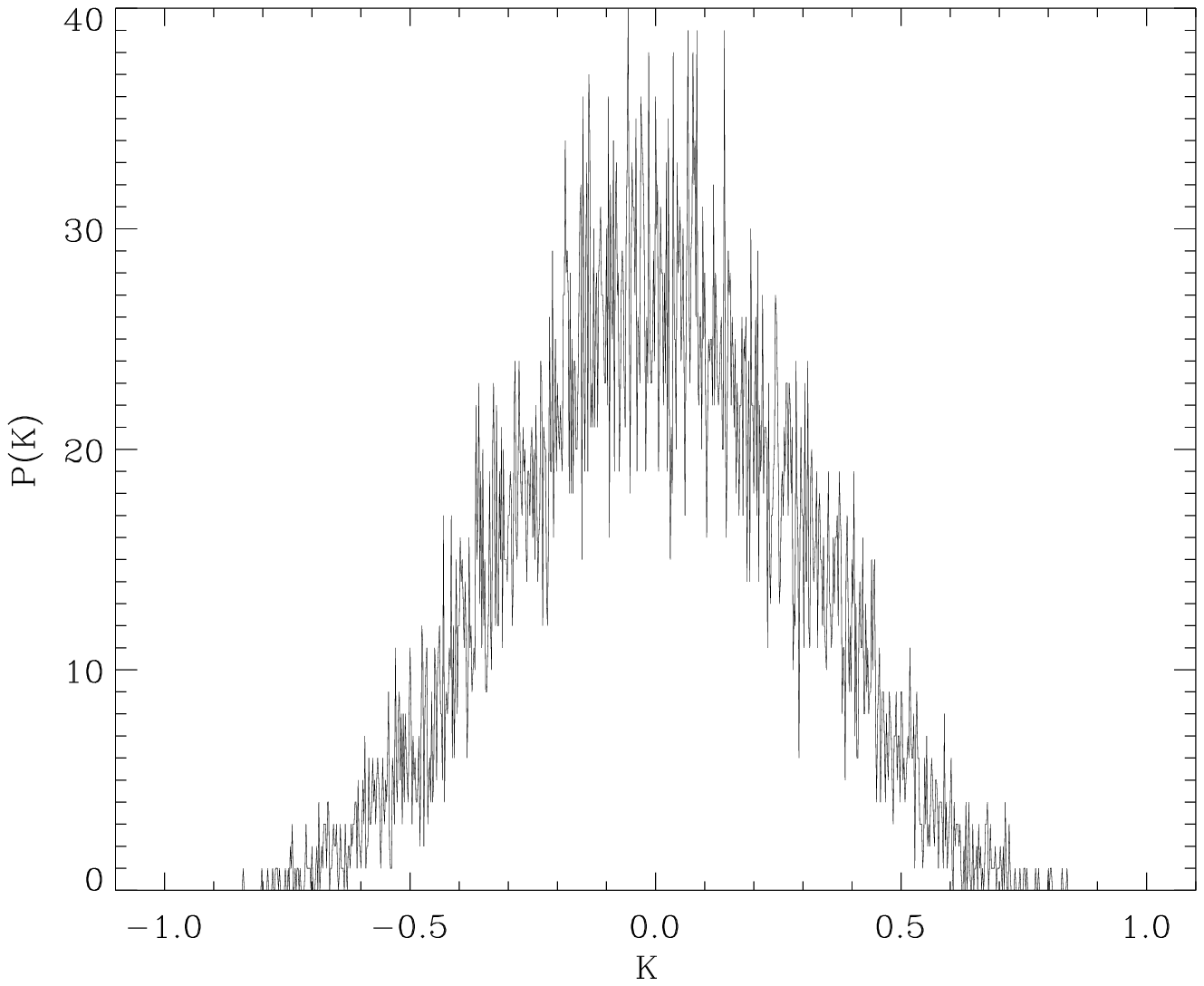,width=5cm}
\psfig{figure=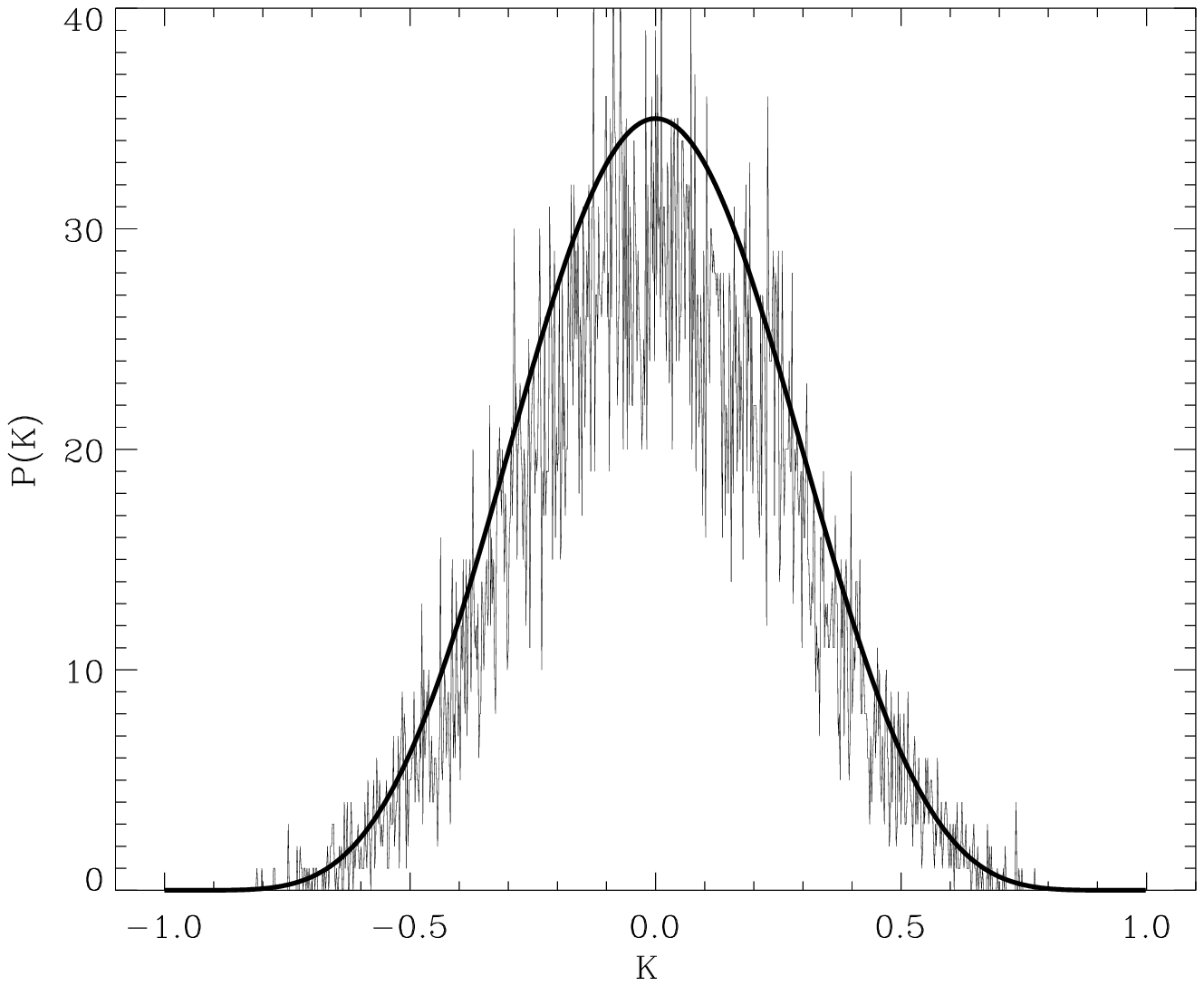,width=5cm}}}
\hbox{\hspace*{-0.5cm}
\centerline{
\psfig{figure=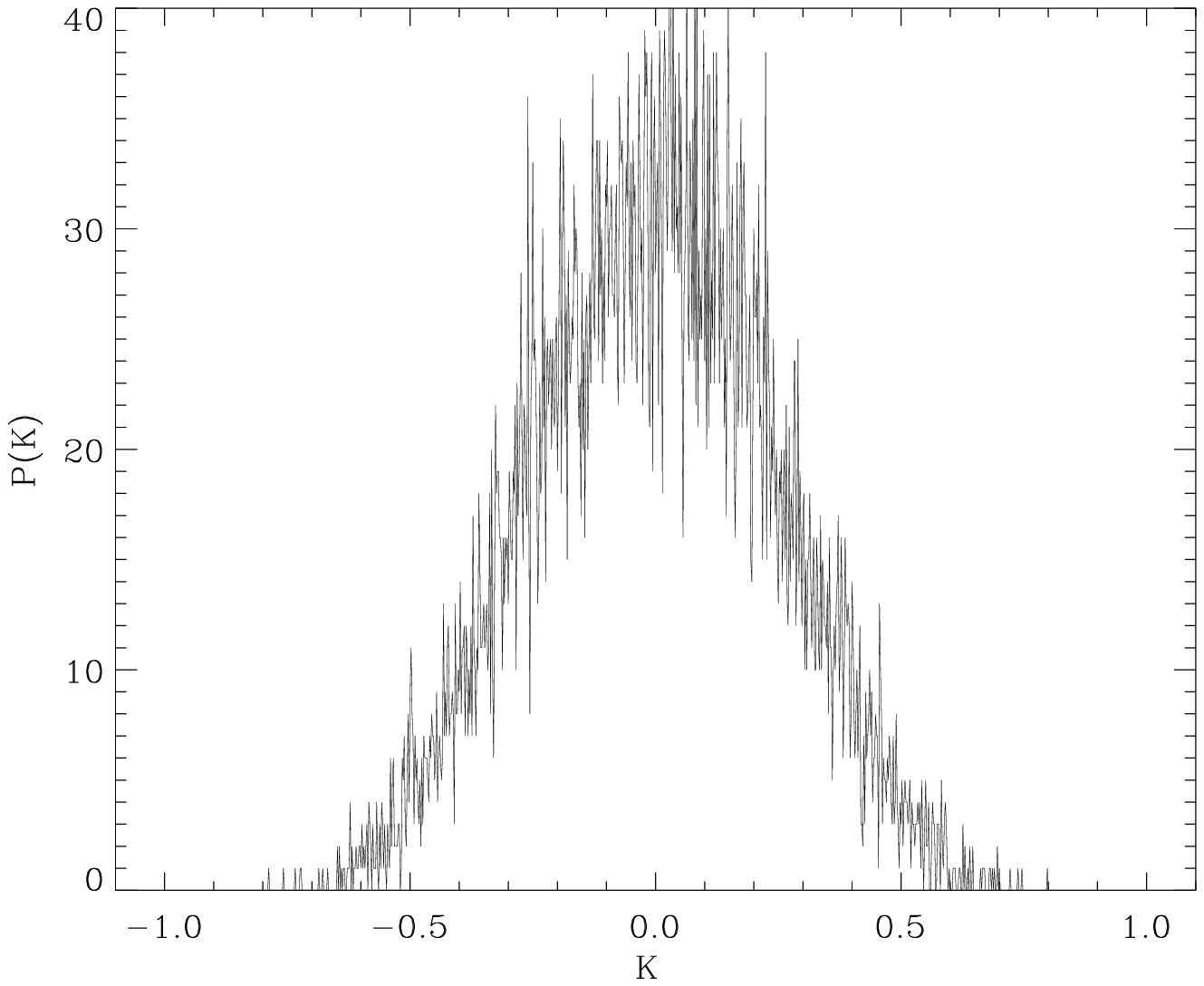,width=5cm}
\psfig{figure=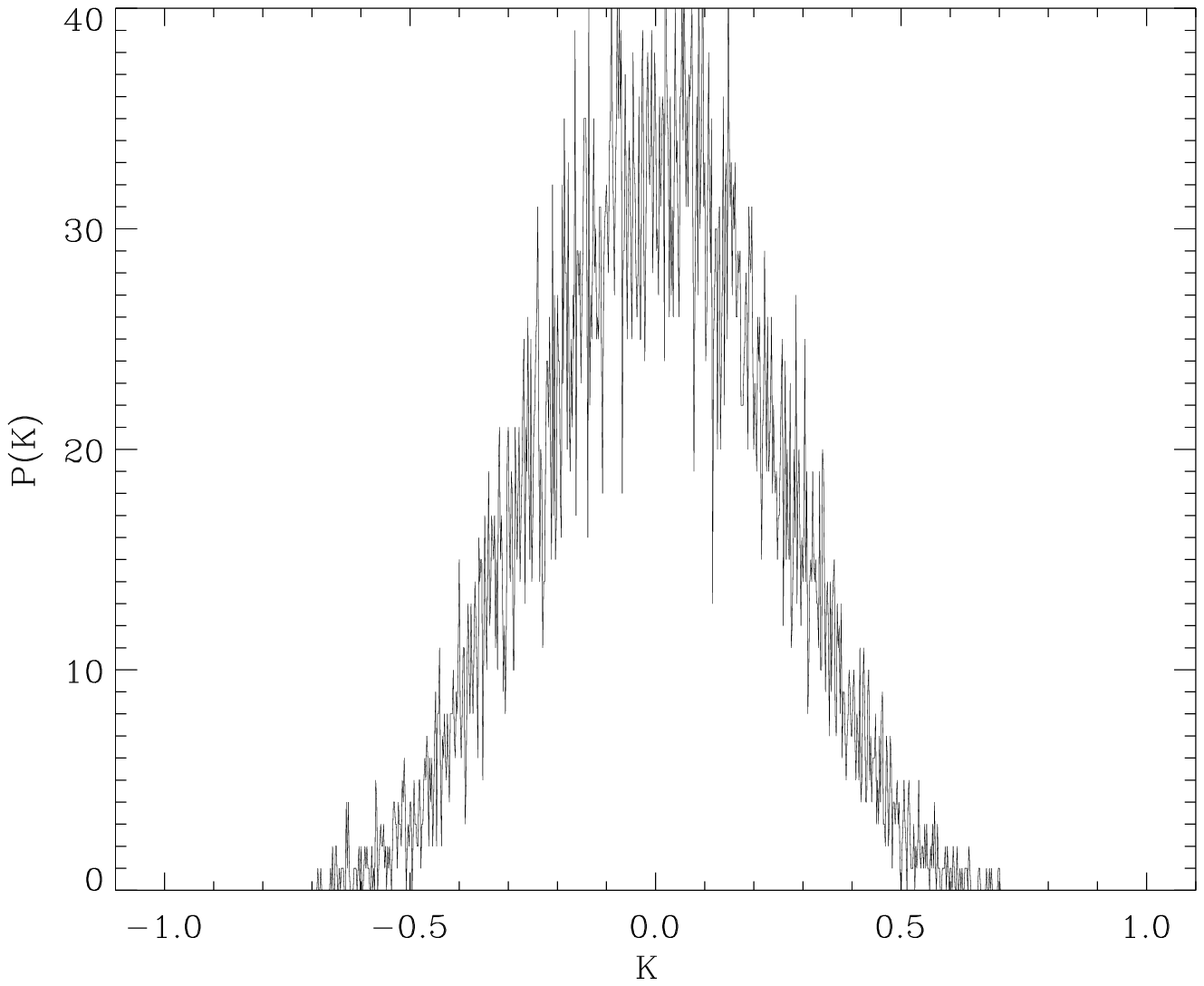,width=5cm}
\psfig{figure=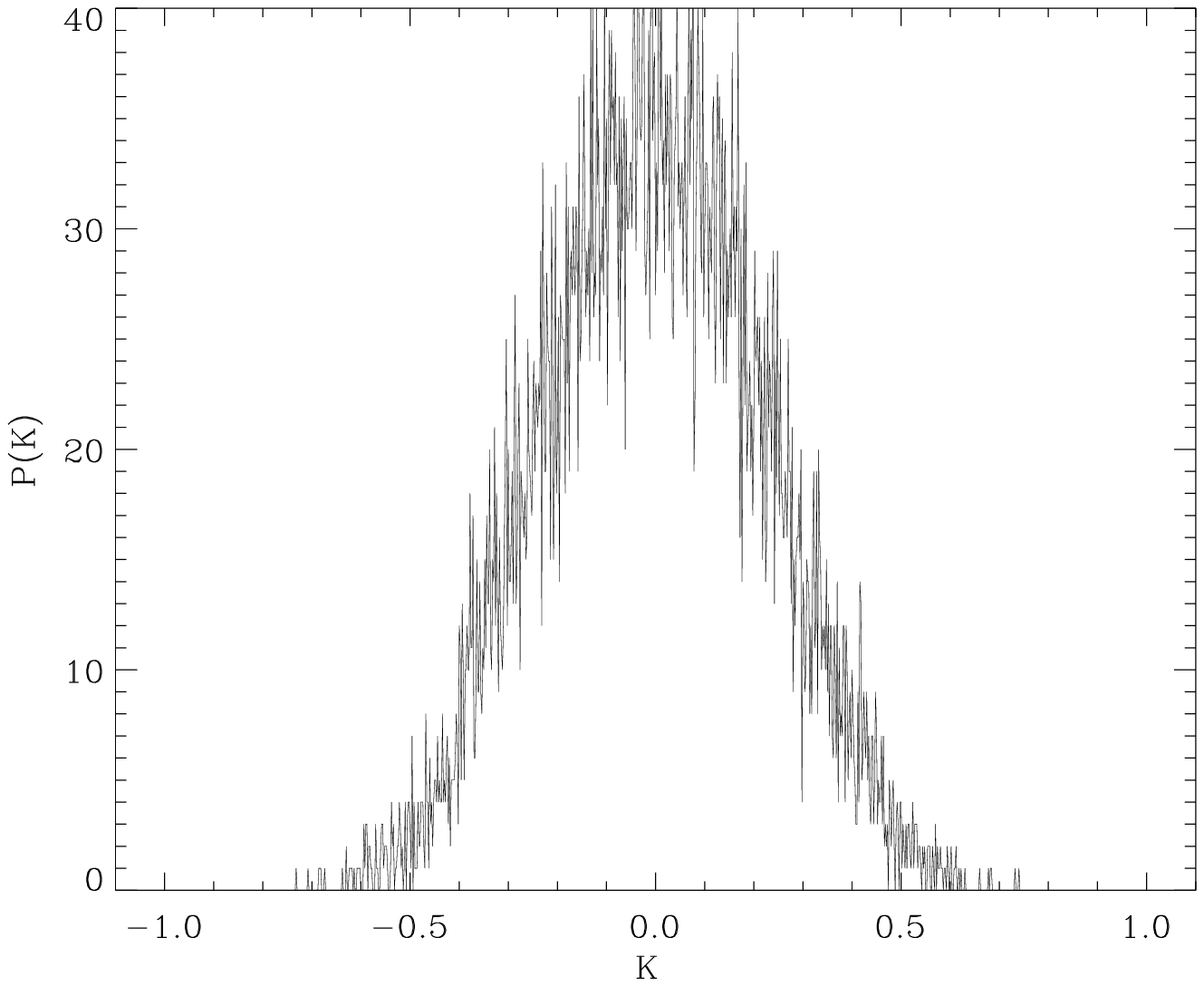,width=5cm}}}
\caption{ The distribution function $P(K)$ for cross--correlation
between random  CMB  and the V-band foreground (left). From the
top left down to the bottom right $\ell=2,3..10$. Note that the
shape of $P(K)$ is the same for any foreground from K--W band
including Haslam et al.
(1982)
signal.}
\label{f1:Verkhodanov_n}
\end{figure*}

In Fig.\,\ref{f1:Verkhodanov_n} we show the probability density
function $P(K_j(\ell))$ versus  $K_j(\ell)$ for the multipoles
$\ell=2\div10$.

For the input  random Gaussian CMB signal  the shape of the distribution
functions  is perfectly fitted by the function
\begin{eqnarray}
P(K,\ell)=A_\ell(1-K^2)^{\ell-1}, \label{p:Verkhodanov_n}
\end{eqnarray}
where $A_\ell$ is the normalization constant. Using $P(K,\ell)$
for the input signal shown in Fig.\,\ref{f1:Verkhodanov_n} (top
left), we have found the first $<K>=-0.00043$ and the second
$<K^2>=0.19934$ moments of $P(K,\ell=2)$, which is in agreement
with corresponding value $<K^2>=0.2$ from
Eq.(\ref{p:Verkhodanov_n}).

\begin{figure*}
\hbox{\hspace*{-0.5cm}
\centerline{\psfig{figure=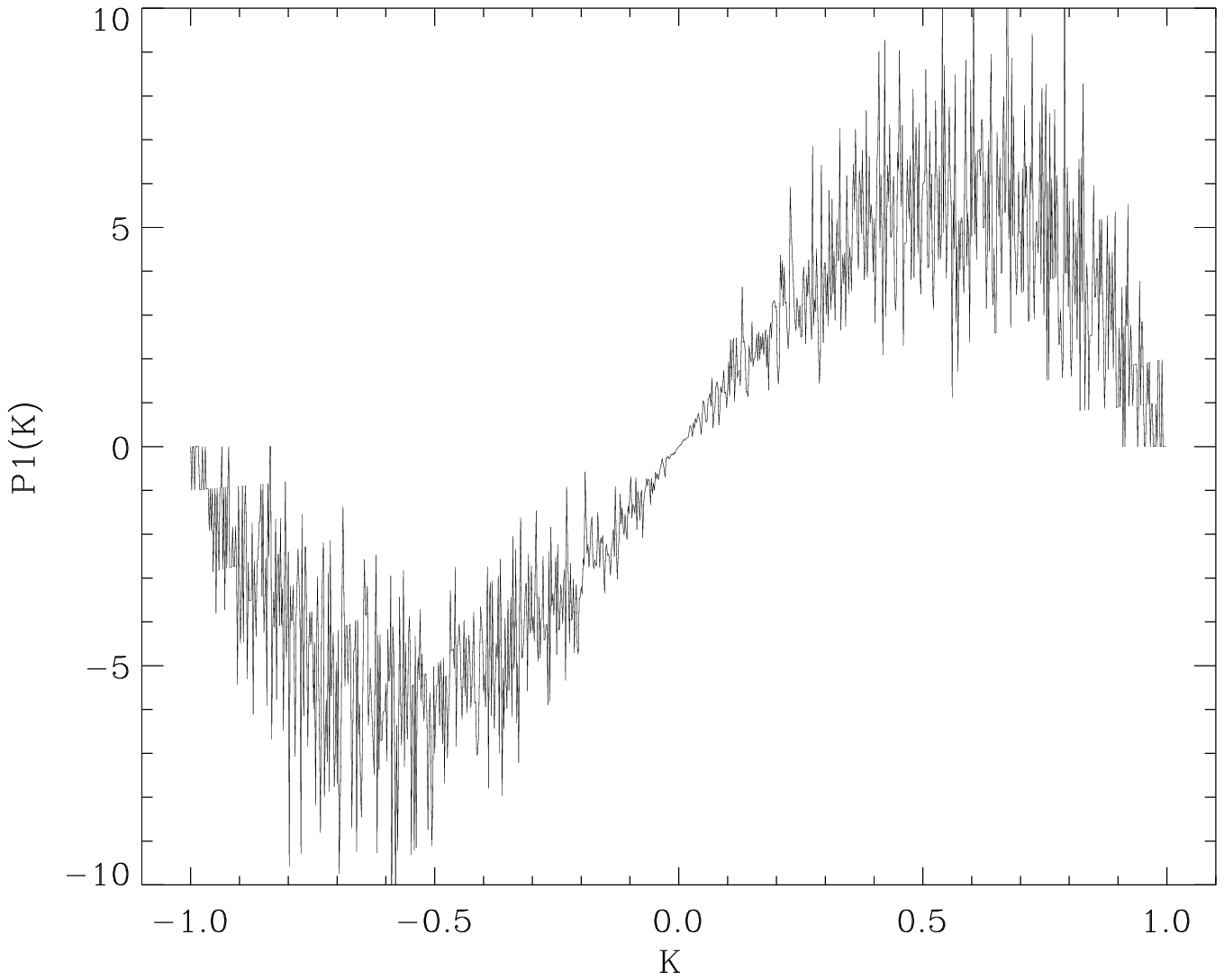,width=7cm}}}
\hbox{\hspace*{-0.5cm}
\centerline{\psfig{figure=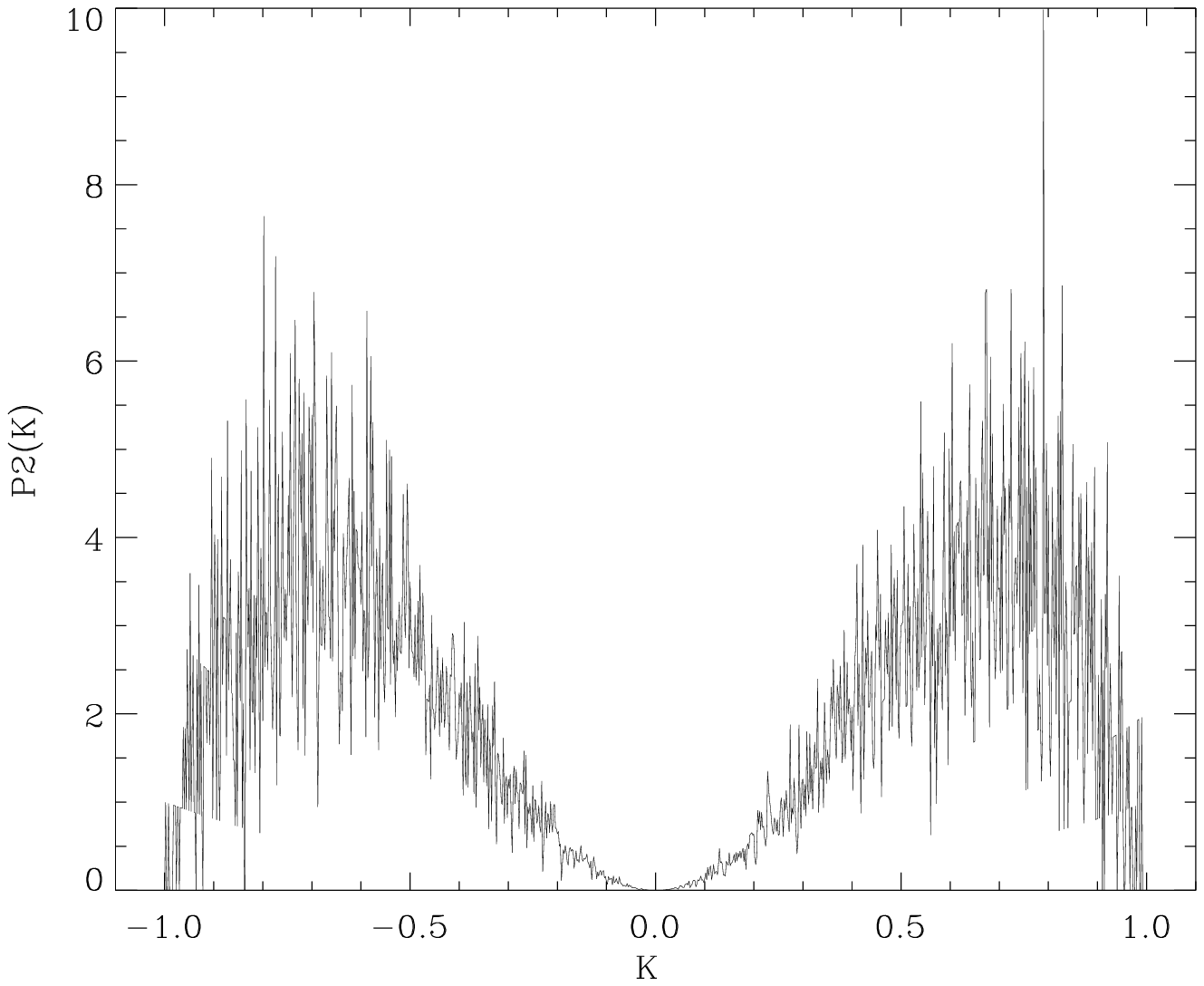,width=7cm}}}
\hbox{\hspace*{-0.5cm}
\centerline{\psfig{figure=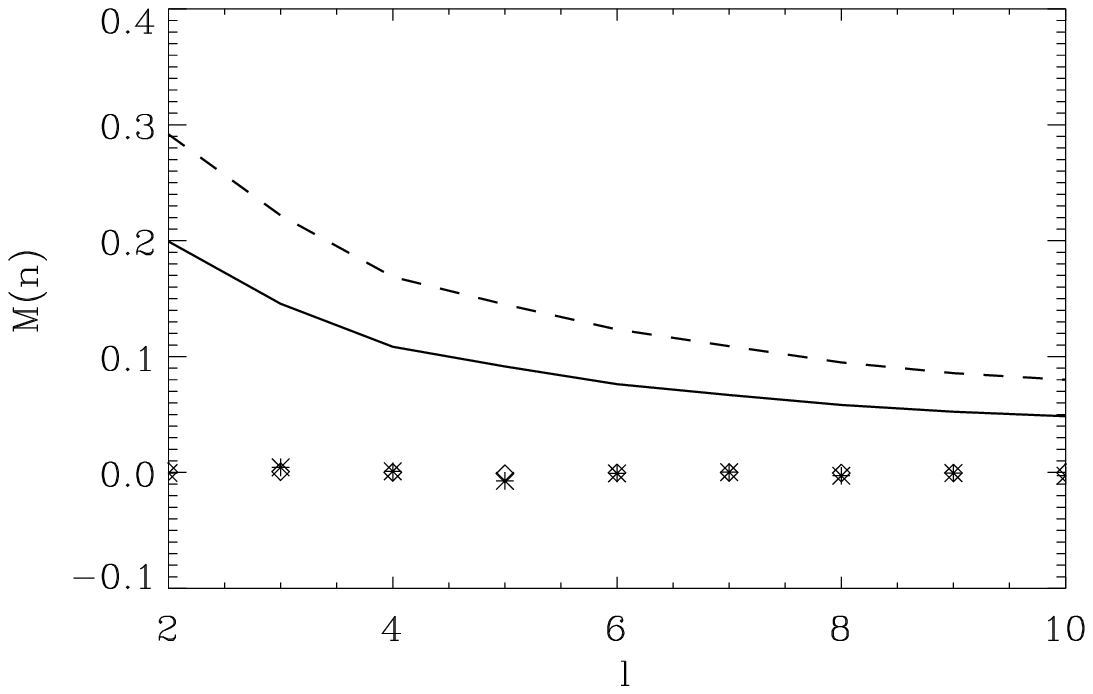,width=7cm}}}
\caption{
The distribution functions $KP(K)$ (top) and $K^2P(K)$ (middle)
for cross--correlation
between random realizations of the CMB signal and
the V band foreground. Bottom plot  shows the distribution of
$\langle K^n(\ell)\rangle$
for $\ell=2-10$. The solid line corresponds to $\langle K^2(\ell)\rangle$,
the dash line is for
$\langle K^4(\ell)\rangle^{\frac{1}{2}}$, the stars are for
$\langle K(\ell)\rangle$, the rhombuses are for
 $\langle K^3(\ell)\rangle$.
} \label{fd1:Verkhodanov_n}
\end{figure*}

\begin{table*}
\caption{The table of $S$ coefficients for 10000 realizations of
the CMB input and output maps. The top row contains data for the
quadrupole component, the bottom one does for the octupole. $\Re$
and $\Im$ mark real and imaginary parts of the $a_{\ell,m}$
coefficients.
}
{\begin{tabular}{c|r|r|c|c|c|c|c} 
\hline
$\ell=2$&$2,0$&$\Re(2,1)$&$\Im(2,1)$&$\Re (2,2) $&$\Im(2,2)$& -&-\\ 
\hline
 $S=-1$ &2148 &361 & 55 &1756 & 257& -&-\\
\hline \hline
$\ell=3$&$(3,0)$& $\Re(3,1)$&$\Im(3,1)$ &$\Re(3,2) $ &$\Im(3,2)$ &$\Re(3,3)$&$\Im(3,3)$\\
\hline
$S=-1$ &232& 1852&337 & 106&367 &533&355 \\  
 \hline
\end{tabular}
\label{Scoef:Verkhodanov_n}
}
\end{table*}

At the end of this section we would like to present the
distribution functions for  the first and second moments of the
coefficient of cross-correlation between the Gaussian realizations
of the CMB signal and the foreground. These moments determine the
bias of the ILC signal and the  power spectrum. It is not
supprising that the first, $KP(K)$ and the second $K^2P(K)$
moments shown in Fig.\,\ref{fd1:Verkhodanov_n}, have two points of
extrema, which can be easily estimated from
Eq.(\ref{p:Verkhodanov_n}):
\begin{equation}
 K^{n,\ell}_{\pm}=\pm\left(\frac{n}{2(\ell-1)+n}\right)^{\frac{1}{2}},
\label{ex:Verkhodanov_n}
\end{equation}
where $n$ is the order of moment $K^nP(K)$, and $\ell$ is the
multipole number. For $\ell=2$ and $n=1$ we have $K^{1,2}=\pm
0.577$ and $K^{2,2}=\pm 0.707$. This result clearly demonstrates
that if the sign of $\kappa$ coefficient is fixed, the most
probable values of $K$ is $K^{1,2}$, and the most probable vale of
$K^2$ is $K^{2,2}$. The bottom plot of Fig.\,\ref{fd1:Verkhodanov_n}
shows the value of the moments \mbox{$M(n)=\langle
K^nP(K,\ell)\rangle$} for the ensemble of realization of the
random Gaussian process. For the second moment $n=2$ the shape of
the function $M(n)$ is $(2\ell+1)^{-1}$, in agreement with
estimation
\cite{saha:Verkhodanov_n} 
of the bias of the power spectrum. From
Eq.(\ref{ex:Verkhodanov_n}) one can find the most probable value
of $K^2$ for the quadrupole component $K^2_{prob}=0.5$. This
means that the most probable value of the factor $(1-K^2_{prob})$ is
0.5, and then from Eq.(\ref{tt11:Verkhodanov_n}), we get
$C_c(\ell=2)=2C_{ilc}(\ell=2)$. At the same time the most probable
value of the bias of the ILC CMB is determined by the parameter
$\kappa=K^{1,2}=\pm 0.577$. One can see that this value is
remarkably close to the WMAP ILC(III) coefficient of the
cross-corelation with the foreground.

\section{ PECULIARITIES OF QUADRUPOLE END OCTUPOLE }

In addition to the cross-correlation with the foreground,
   we have discovered one more  feature of the ILC method,
which could have a significant impact on solution of the problem
of the ILC(III) \mbox{quadrupole \cite{nv_quad:Verkhodanov_n}.} We
call this effect as a flip of the sign of the (2,0) component of
the ILC quadrupole with respect to the sign of the true CMB (2,0)
component. To understand quantitatively this effect, let us come
back to Eqs.(\ref{tt9:Verkhodanov_n}) and
(\ref{tt10:Verkhodanov_n}) for the quadrupole component
$\ell=2$ and take under consideration the $(2,m=0)$ mode:
\begin{eqnarray}
\nonumber  |c^{ilc}_{2,0}|\cos(\xi_{2,0})=|c_{2,0}|\cos(\eta_{2,0})\\
 -\kappa\left[\frac{\sigma_c}{\sigma_F}\right]^{\frac{1}{2}}
            |F_{2,0}|\cos(\phi_{2,0}), \hspace{0.6cm}
 \label{flip1:Verkhodanov_n}
\end{eqnarray}
where $|..|$ denotes the modulus, and $\xi_{2,0},\eta_{2,0}$ and
$\phi_{2,0}$ are the phases of $2,0$ component of the ILC
quadrupole, the true quadrupole, and the foreground
correspondingly. Since the $m=0$
 mode has the only real part, these phases simply mean the sign of the (2,0)
components. For our further analysis it is important to note that
for all K--W foregrounds $|F_{2,0}|\gg |F_{2,1}|,|F_{2,2}|$ and
$\phi_{2,0}=\pi$. Taking into account
Eq.(\ref{tt10:Verkhodanov_n}), one can easily find that

\begin{eqnarray}
 \kappa\simeq -\frac{|c_{2,0}||F_{2,0}|\cos(\eta_{2,0})}
{\sigma_c\sigma_F}(1-\varepsilon), \nonumber\\
\varepsilon=\frac{2\sum_{m=1}^2|c_{2,m}||F_{2,m}|
\cos(\eta_{2,m}-\phi_{2,m})}{\cos(\eta_{2,0})|c_{2,0}||F_{2,0}|},
\label{flip2:Verkhodanov_n}
\end{eqnarray}
where $\eta_{2,m},\phi_{2,m}$ are the phases of the true CMB and
the foreground with $m=1$ and $m=2$ components of the quadrupole. Thus,
after substitution of $\kappa$ from Eq.(\ref{flip2:Verkhodanov_n})
into Eq.(\ref{flip1:Verkhodanov_n}), we get

\begin{eqnarray}
|c^{ilc}_{2,0}|\cos(\xi_{2,0})=|c_{2,0}|\cos(\eta_{2,0})
             \frac{\mu+\varepsilon}{1+\mu},
\label{flip3:Verkhodanov_n}
\end{eqnarray}
where $$\mu=(|F_{2,1}|^2+|F_{2,2}|^2)/|F_{2,0}|^2\ll 1.$$ From
Eq.~~(\ref{flip3:Verkhodanov_n}) one~~~ can ~~get ~~that~~ if
~~~$\varepsilon \ll \mu$,~~~ then \mbox{$\xi_{2,0}=\eta_{2,0}$}~~
and ~~$|c^{ilc}_{2,0}|=\mu|c_{2,0}|\ll |c_{2,0}|$. Thus, the ILC
$|c^{ilc}_{2,0}|$ mode  is smaller than corresponding $|c_{2,0}|$
by the factor $\sim \mu$. However, if $\varepsilon \gg \mu$, we
have
\begin{eqnarray}
\nonumber |c^{ilc}_{2,0}|\cos(\xi_{2,0}) \hspace{4.3cm}
\\\simeq 2\frac{\sum_{m=1}^2|c_{2,m}|
              |F_{2,m}|\cos(\eta_{2,m}-\phi_{2,m})}{|F_{2,0}|}
 \label{flip4:Verkhodanov_n}
\end{eqnarray}
and now the phase $\cos(\xi_{2,0})$ is determined by the sign of
the nominator of the right part
in Eq.(\ref{flip4:Verkhodanov_n}). This result tells
us that for some of realizations of the random CMB, the sign of
the input signal is no matter. The reconstructed ILC phase
$\xi_{2,0}$ can be  the same $(\eta_{2,0})$, or opposite $
\eta_{2,0}\pm\pi$ to the phase of the true CMB.
\begin{figure*}
\hbox{\hspace*{-0.5cm}
\centerline{\psfig{figure=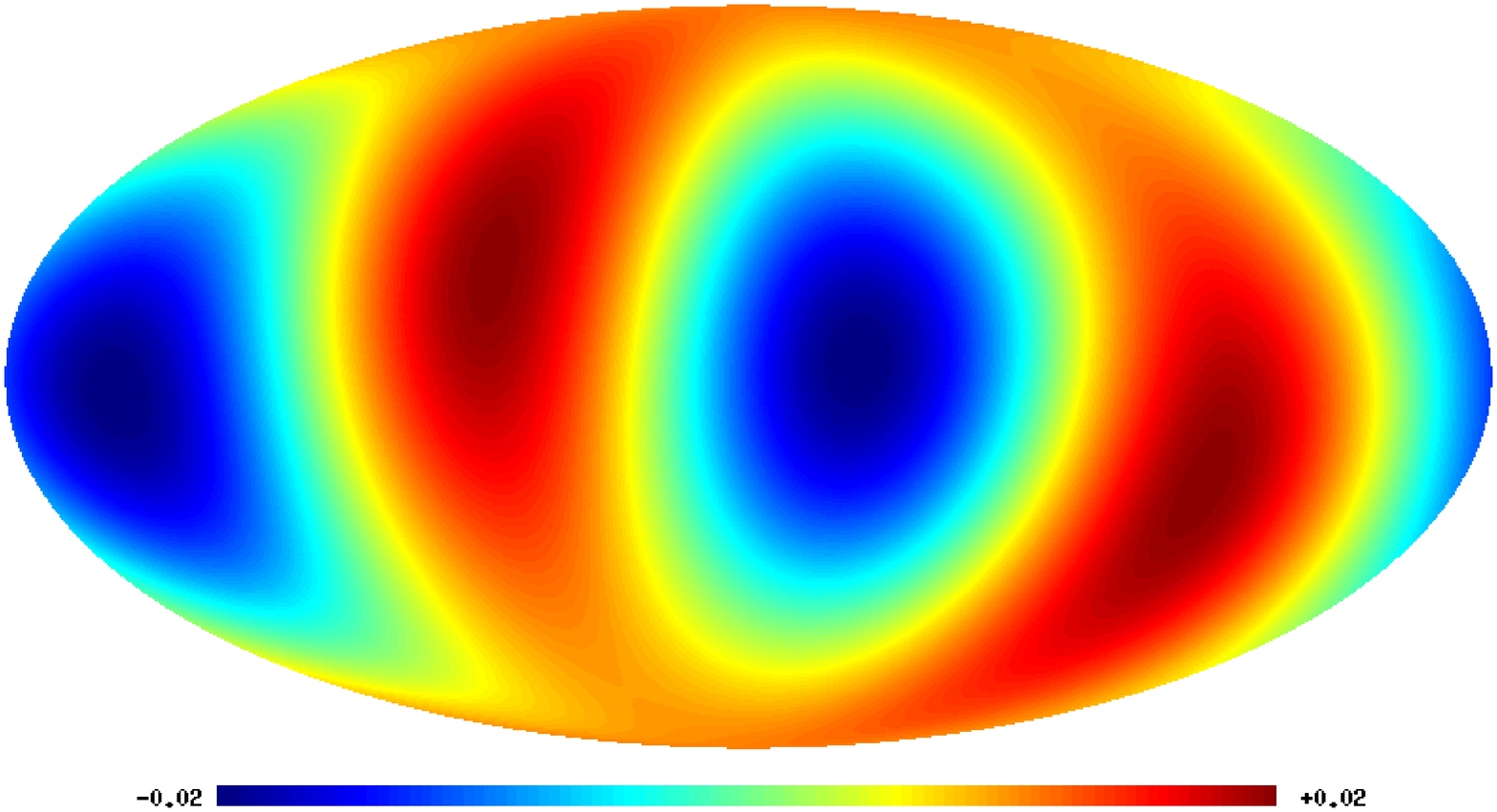,width=6cm}}}
\hbox{\hspace*{-0.5cm}
\centerline{\psfig{figure=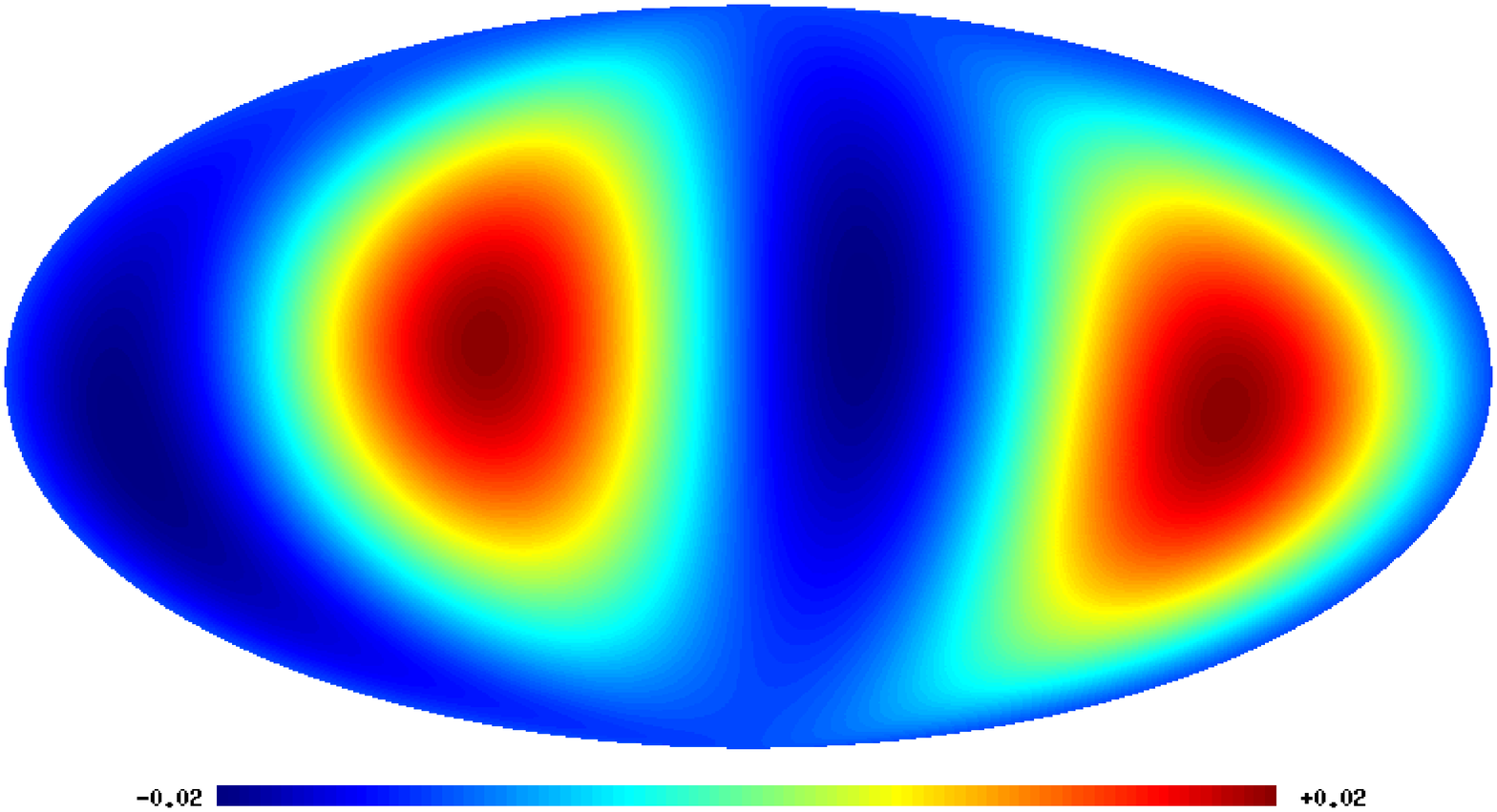,width=6cm}}}
\hbox{\hspace*{-0.5cm}
\centerline{\psfig{figure=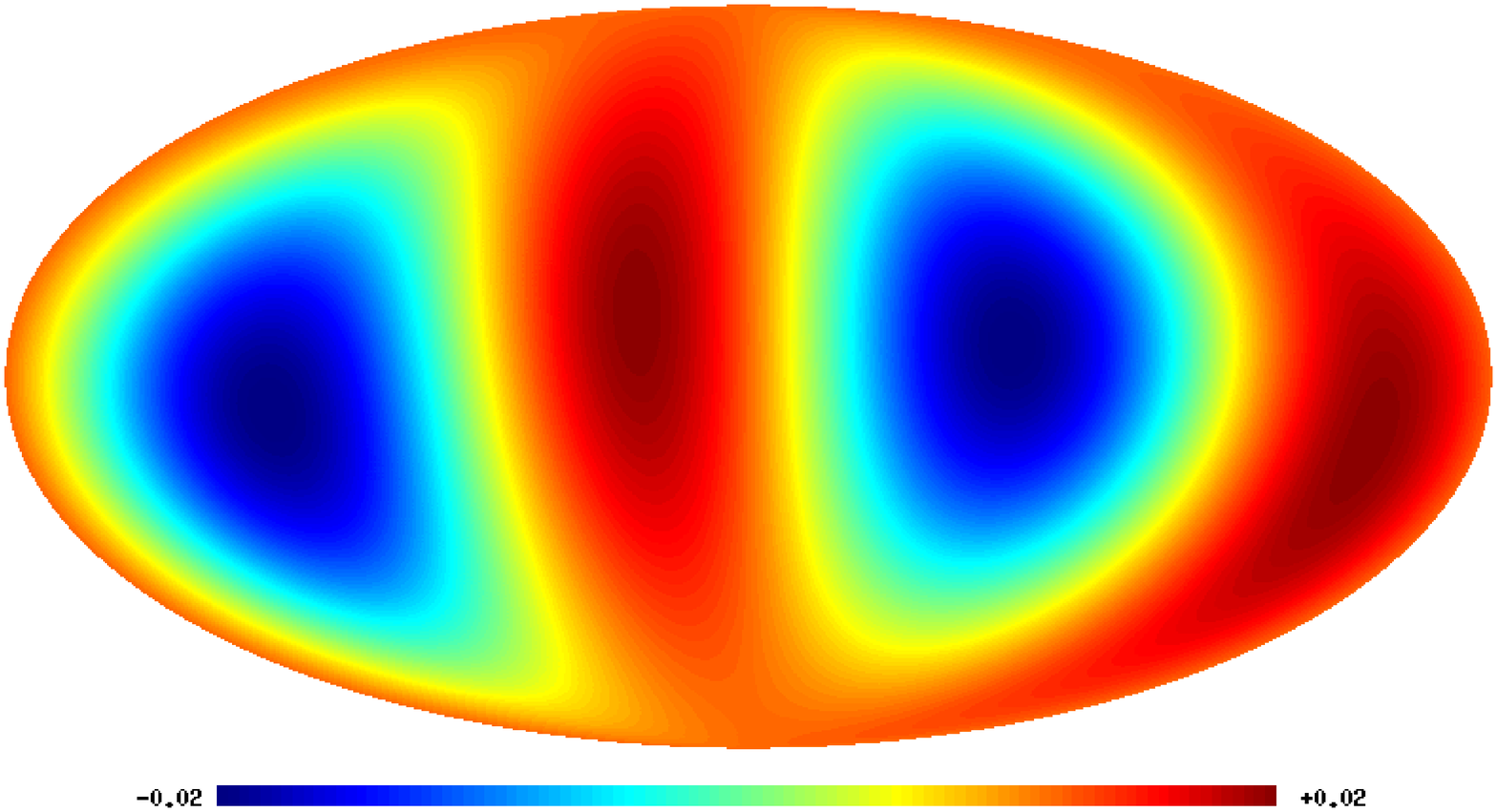,width=6cm}}}
\hbox{\hspace*{-0.5cm}
\centerline{\psfig{figure=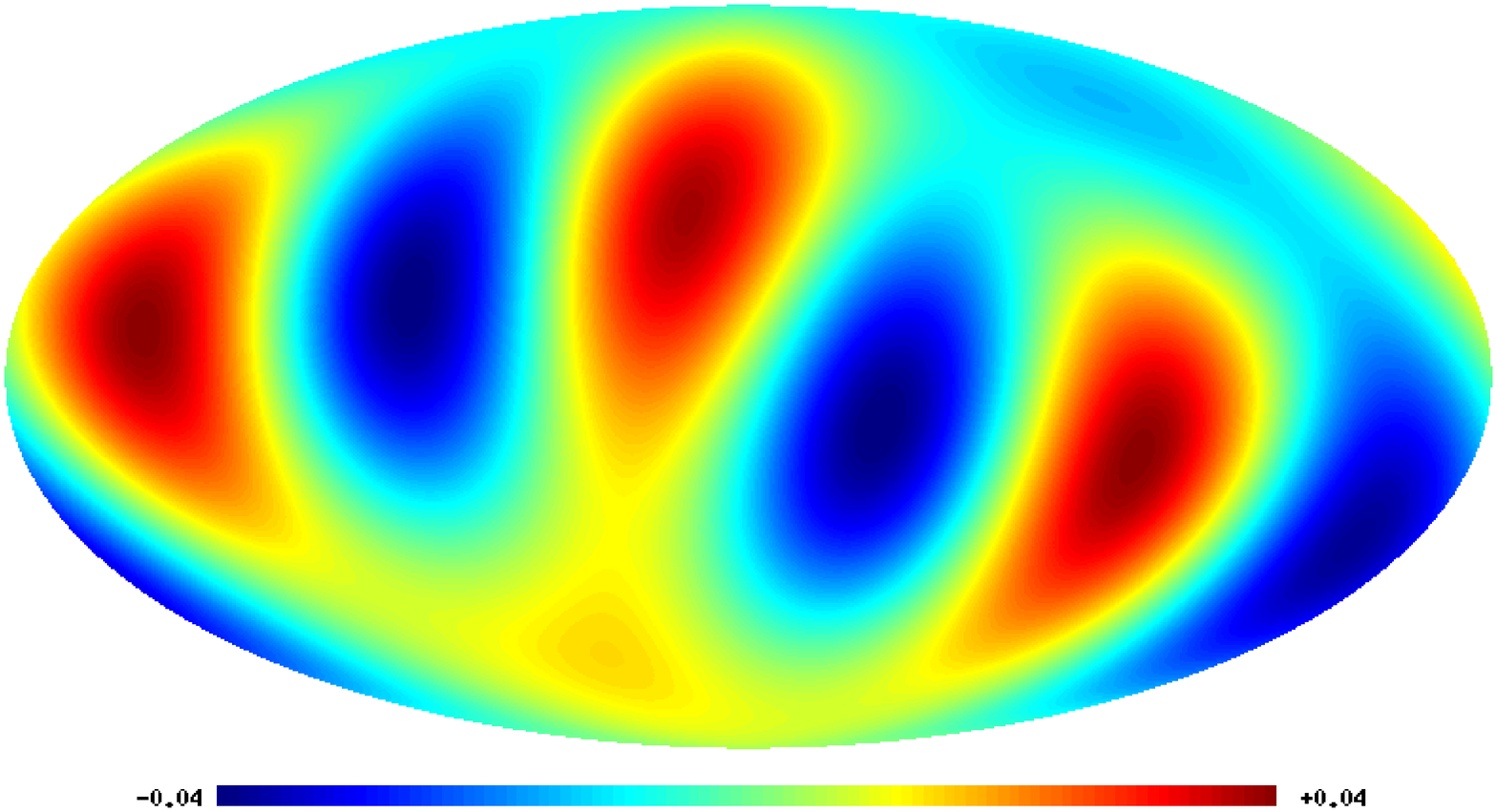,width=6cm}}}
\hbox{\hspace*{-0.5cm}
\centerline{\psfig{figure=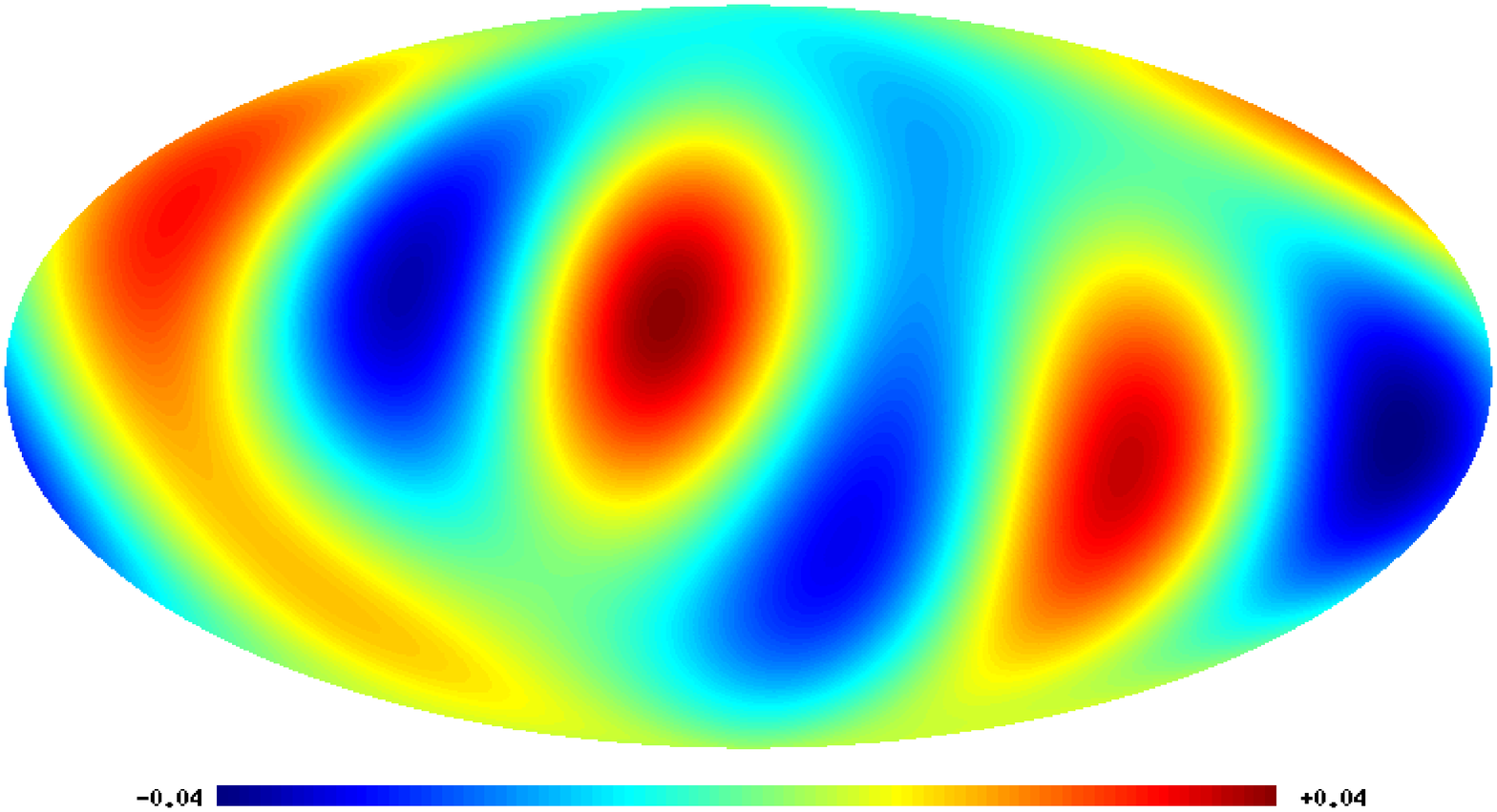,width=6cm}}}
\caption{ The illustration of the flip-effect for the ILC(III)
quadrupole and octupole. Top  is the ILC(III) quadrupole, the
second from top  is the ILC(III) quadrupole with flip-effect of
the (2,0) component, the third from the top is the quadrupole with
flip-effect of the (2,2) component. The second from the bottom and
the bottom panels show  the ILC(III) octupole and the flip-effect
for the (3,1) octupole component. }
\label{ffs:Verkhodanov_n}
\end{figure*}

To show that this effect takes place in numerical simulations of
the ILC method,
 we  consider the  10000
realization input and output maps for the $c_{2,0}$ quadrupole component,
presented in
\cite{eriksen2004a:Verkhodanov_n}. 
As was pointed out in
\cite{wmap3temp:Verkhodanov_n}, 
the difference between the WMAP and Eriksen et al. approach
\cite{eriksen2004a:Verkhodanov_n} for derived CMB signals is
caused by different choice of disjoint regions in the Galactic
plane area, rather than the different bias.
 Using an estimator
$$S=s^{in}_{2,0}\cdot s^{out}_{2,0}=\cos\eta^{in}_{2,0}\cdot\cos\xi^{out}_{2,0},$$
where $s^{in, out}_{2,0}=+1$~ or $-1$ for the positive and negative sign of
$c^{in,out}_{2,0}$ components, correspondingly,
we  have found that for 2148 realizations $S=-1$.
Moreover, since for the foregrounds
$s^f_{2,0}=-1$ for all K--W the WMAP bands,
practically $43\%$
of the realizations having $s^{in}_{2,0}=-1$ after using the ILC method
change the sign to  $s^{out}_{2,0}=1$ (the effect of flip).
We extend our analysis to the octupole component of
the Eriksen et al.
\cite{eriksen2004a:Verkhodanov_n}
ensemble of input and output signals and
have found that this effect still takes place, but the number of events
is slightly smaller than one for the quadrupole component.

In Table 2 we present the number of events for $S=-1$, when input and
output components of the quadrupole and octupole have the different sign.

One can see from this Table 2, that for the $(2,0)$ and $(3,1)$
components the number of events with $S=-1$ reaches the maximum
($2148$ and $1852$ events, correspondingly). To show that the
flipping of the sign of the $(2,0)$ component can be responsible
for increasing of the CMB-foreground cross-correlation, we take
one of the realizations of input \mbox{Eriksen et al.}
\cite{eriksen2004a:Verkhodanov_n}
maps, namely \mbox{``in--00008''}, and calculated the
$K^{in}(\ell=2)$ coefficient for this map and the Haslam et al.
\cite{haslam:Verkhodanov_n}
synchrotron map. We have  $K^{in}(\ell=2)= -0.222$. For
this input map, \mbox{$c^{in}_{2,0}=-9.944\mu K$.} For the output
CMB maps, named as ``out--00008'',
\mbox{$c^{out}_{2,0}=11.457\mu K$} and \mbox{$K^{out}(\ell=2)= -0.522$}.
Then, we change the sign of $c^{out}_{2,0}$
 component and recalculate the coefficient of
the cross-correlation once again. We have
\mbox{$K^{out}(\ell=2)=-0.2466$,} which is practically the same
for the input map.
This example clearly demonstrates that for input realizations mentioned above
with negative $c^{in}_{2,0}$,
this component of the signal would be quite likely reconstructed with
the opposite sign which leads to increasing a negative level of
cross-correlation coefficients $K^{out}(\ell=2)$. Roughly
speaking, the value of the coefficients $K$ depends on the
flip-effect of the sign for the (2,0) component.

To show corresponding changes of the images of the quadrupole and
octupole related to the flip-effect, in
Fig.\,\ref{ffs:Verkhodanov_n} we show the ILC(III) quadrupole and
octupole with the different sign of the (2,0), (2,2) and (3,1)
components. As one can see the flip-effect changes significantly
the morphology of the maps. This is not surprise since all these
$\ell+m$ even components are the most powerful components of the
ILC(III) signal.

\section{ CONCLUSION }

We have investigated the cross--correlation of the ILC(I) and
the ILC(III) with the WMAP MEM
 foregrounds,
and shown that  these correlations tightly coupled to the bias of
the ILC CMB signal. By using Monte-Carlo simulations we  found the
probability distribution function for the coefficient of
cross-correlation between the true CMB and the foreground signals.
For debiasing of the ILC  CMB, we need to know exact value of the
coefficient $K(\ell)$. Since the ILC MV method does not provide
any additional information about the bias, any assumptions about
the properties of the foregrounds should  leave corresponding
residuals on the debiased ILC signal and its power spectrum (see
for the comparison \cite{wmap3temp:Verkhodanov_n,park3y:Verkhodanov_n,saha:Verkhodanov_n}). These
uncertainties reflect directly our ignorance of the exact value of
the realization of the random process $K(\ell)$, making the
reconstruction of the CMB unstable.

There is another reason for instability of reconstruction of
the quadrupole and the octupole of the CMB related to
the CMB-foreground bias. We have discovered the flip-effect of the sign
for even $\ell +m$ modes
((2,0) and (2,2) components
of the quadrupole and the (3,1) component of the  octupole).
The corresponding probability
of this effect is about 20\% (see  Table 1). This effect could
have significant impact on the debiasing technique.


\begin{acknowledgements}

We acknowledge the use of the NASA Legacy Archive for extracting
the \wmap data. We also acknowledge the use of \healpix
\footnote{\tt http://www.eso.org/science/healpix/} package
\cite{healpix:Verkhodanov_n}              
to produce $\alm$ from the \wmap data and the use of
\glesp \footnote{\tt http://www.glesp.nbi.dk} package
\cite{glesp:Verkhodanov_n,glesp2:Verkhodanov_n}.       
OV thanks RFBR for partly supporting by the grant No\,08-02-00159.

\end{acknowledgements}

\newcommand{\autetal}[2]{{#1,\ #2. \etal,}}
\newcommand{\aut}[2]{{#1,\ #2.,}}
\newcommand{\saut}[2]{{#1,\ #2.,}}
\newcommand{\laut}[2]{{#1,\ #2.,}}

%
%
\newcommand{\refs}[6]{#5, #2, #3, #4} 
\newcommand{\arefs}[7]{#5, #2, #3, #4 (#6)} 
\newcommand{\unrefs}[6]{#5, #2 #3 #4 (#6)}  

\newcommand{\book}[6]{#5, {\it #1}, #2} 
%
\newcommand{\proceeding}[6]{#5, in #3, #4, #2} 

\newcommand{\combib}[3]{\bibitem[\protect\citename{#1 }#2]{#3:Verkhodanov_n}} 

%
%
\def\nature{nat}
\def\prl{Phys.\ Rev.\ Lett.}
\def\prd{Phys.\ Rev.\ D}
\def\pr{Phys.\ Rep.}
\def\ijmpd{Int. J. Mod. Phys. D}
\def\jcap{J. Cosmo. Astropart. Phys.}
\def\ab{Astrophys. Bull.}
\def\bsao{Bull. SAO}

\def\cup{Cambridge University Press, Cambridge, UK}
\def\princetonpress{Princeton University Press}
\def\worldpress{World Scientific, Singapore}
\def\oxfordpress{Oxford University Press}

\end{document}